\documentclass{clv3}

\usepackage[utf8]{inputenc}
\usepackage{hyperref}
\usepackage{xcolor}
\usepackage{graphicx}
\usepackage{graphicx, caption}
\usepackage{subcaption}
\usepackage{enumitem}
\usepackage{multirow}
\usepackage{amsmath}
\usepackage{hhline}
\definecolor{darkblue}{rgb}{0, 0, 0.5}
\definecolor{black}{rgb}{0,0,0}
\hypersetup{colorlinks=true,citecolor=black, linkcolor=black, urlcolor=black}

\setcitestyle{citesep={,}}

\begin{document}

\title{Harbsafe-162 – A Domain-Specific Data Set for the Intrinsic Evaluation of Semantic Representations for Terminological Data}

\author{Susanne Arndt, M. A.\thanks{Technische Universität Braunschweig, Institut für Verkehrssicherheit und Automatisierungstechnik, Hermann-Blenk-Straße 42, 38108 Braunschweig, Germany. E-mail: s.arndt@tu-braunschweig.de.}}
\affil{Technische Universität Braunschweig}

\author{Dieter Schnäpp, M. A.\thanks{Technische Universität Braunschweig, Institut für Verkehrssicherheit und Automatisierungstechnik, Hermann-Blenk-Straße 42, 38108 Braunschweig, Germany. E-mail: d.schnaepp@tu-braunschweig.de.}}
\affil{Technische Universität Braunschweig}

\maketitle

\begin{abstract}
The article presents Harbsafe-162, a domain-specific data set for evaluating distributional semantic models. It originates from a cooperation by Technische Universität Braunschweig and the German Commission for Electrical, Electronic \& Information Technologies of DIN and VDE, the Harbsafe project. One objective of the project is to apply distributional semantic models to terminological entries, that is, complex lexical data comprising of at least one or several terms, term phrases and a definition. This application is needed to solve a more complex problem: the harmonization of terminologies of standards and standards bodies (i.e. resolution of doublettes and inconsistencies). Due to a lack of evaluation data sets for terminological entries, the creation of Harbsafe-162 was a necessary step towards harmonization assistance. Harbsafe-162 covers data from nine electrotechnical standards in the domain of functional safety, IT security, and dependability. An intrinsic evaluation method in the form of a similarity rating task has been applied in which two linguists and three domain experts from standardization participated. The data set is used to evaluate a specific implementation of an established sentence embedding model. This implementation proves to be satisfactory for the domain-specific data so that further implementations for harmonization assistance may be brought forward by the project. Considering recent criticism on intrinsic evaluation methods, the article concludes with an evaluation of Harbsafe-162 and joins a more general discussion about the nature of similarity rating tasks. Harbsafe-162 has been made available for the community.
\end{abstract}

\section{Motivation}
\label{sec-1}
\textbf{Terminologies}, the “set[s] of designations […] belonging to one special language […]” \cite{1087-1.2000} are a challenging object of standardization. Phenomena like ambiguity and synonymy potentially complicate domain-specific communication\footnote{\textit{Domain} in this sense refers to scientific and technical subjects, not domains of discourse like tweeting.}  \cite[cf.][]{Drewer.2017}. To avoid misunderstandings, the thorough documentation and evaluation of terminologies is a common practice. This process of “systematic collection, description, processing and presentation of concepts […] and their designations […]” \cite{1087-1.2000} is called \textbf{terminology work} and adheres to principles formulated by the discipline of \textbf{terminology}. \textbf{Terminology standardization} as defined by ISO is concerned with the “development of terminological entries […] in terminology standards […] and of terminology sections in other standards, and their approval by a standardizing body […]”\cite{10241-2.2012}. A \textbf{terminological entry} is a “part of a terminological data collection […] which contains the terminological data […] related to one concept” \cite{1087-1.2000}.

Standardizing bodies should only approve terminological entries that are unique and consistent with other standardized terminological entries. Uniqueness  implies that if a subcommittee of a specific technical committee (TC) has already defined a concept within a specific standard, no other subcommittee should standardize it again \cite{820-2.2012}. Terminology standardization should thus result in recommendations for the unambiguous use of terms and subject-specific concepts. As such, it needs to adhere to several requirements: \ref{req1} – \ref{req3} are requirements for concepts and their representations while \ref{req4} and \ref{req5} are requirements for parts of terminological entries.

\begin{enumerate}[label={(\arabic*)}]
\item Each concept shall be described by only one terminological entry. \label{req1}
\item Each concept shall be described by only one definition. \label{req2}
\item Each concept shall be represented by only one preferred designation.\label{req3}\footnote{Whether \ref{req3} and \ref{req5} can be considered hard and fast “requirements” is in fact a matter of debate. Terminology standards like ISO 10241-1:2011 propagate that “every effort shall be made to avoid use of a single term for multiple concepts and multiple terms for a single concept” \cite{10241-1.2011}. This partially corresponds to our requirements \ref{req3} and \ref{req5}. The modal \textit{shall} makes this a requirement for standards that needs to be “fulfilled and from which no deviation is permitted if compliance with the document is to be claimed” \cite{ISO.2016}. In fact, it is even stricter than our requirements \ref{req3} and \ref{req5}: Each concept shall be represented by \emph{only} one designation and each designation shall represent \emph{only} one concept. While this stricter requirement is still found obligatory in the discipline of terminology science \cite[cf.][]{Drewer.2017}, there are discussions of its limitations \cite[cf.][]{Roelcke.2004}. Accordingly, ISO 10241-1 \cite{10241-1.2011} or DIN 820-2:2018-09 \cite{820-2.2018} need to cope with the fact that natural languages never follow this strict demand.}
\item Each terminological entry shall have only one definition. \label{req4}
\item Each terminological entry shall have only one preferred designation. \label{req5}
\end{enumerate}

Unfortunately, these requirements are not always met. Even standards bodies may produce terminologies that show ambiguity, synonymy, and even inconsistencies \cite[cf.][]{Arndt.2015}. The reality for requirements \ref{req1}, \ref{req2}, and \ref{req3}, is thus better depicted by \ref{req1'}, \ref{req2'}, and \ref{req3'}.

\begin{enumerate}[label={(\arabic*')}]
\item A concept is potentially described by more than one terminological entry. \label{req1'}
\item A concept is potentially described by more than one definition. \label{req2'}
\item A concept is potentially represented by one or more preferred designations.\label{req3'}\footnote{These are of course not the only flaws of a terminology: there could also be gaps \cite[cf.][]{Muller.2015}: (a) a concept is described by no terminological entry, (b) a concept is described by no definition, (c) the preferred designation of a concept may not have been chosen. This article, however, does not address such issues, since it is concerned with terminological data that exist (in a terminological resource) and not with those that do not.}
\end{enumerate}

In other words, the whole set of standardized terminology may contain non-unique terminological entries called doublettes in the context of terminological databases. A \textbf{doublette} is defined as a “terminological entry […] that describes the same concept as another entry” \cite{26162.2012} and is therefore applicable to terminological entries in standards as well. Requirements \ref{req1}, \ref{req2}, and \ref{req3} can thus be summarized by requirement \ref{req6}.

\begin{enumerate}[label={(\arabic*)}]
\setcounter{enumi}{5}
\item A set of terminological entries shall not contain doublettes. \label{req6} 
\end{enumerate}

Requirements \ref{req4} and \ref{req5} are met comparatively easily. It should practically never happen that a terminological entry has more or less than one definition (for the same language).\footnote{This might be the case for work in progress when a technical committee has not decided yet which definition is most appropriate.} Still, it may occur that a definition does not satisfy other quality criteria as stated by \citet{1087-1.2000}. As regards requirement \ref{req5}, it is uncommon that more than one designation is marked as preferred within one terminological entry. In addition, it is common practice to set the only term in a terminological entry as the preferred one automatically. Whenever an entry has a preferred term, the reasons for the choice of this preferred designation might be opaque. In standardization, style guides for standards, for example DIN 820-2:2018-09 \textit{Standardization - Part 2: Presentation of documents}, regulate the internal structure and information types a terminological entry consists of. Such requirements for the formal design of terminological entries are usually thoroughly checked before publication.

There are several potential reasons for doublettes. First, there are many participants involved in standardization activities, all of them coming from different backgrounds and pursuing different goals. Second, standardization activities may run in parallel so that several descriptions of the same concept may result. Since there are no governing bodies that have the responsibility, authority or the means to check all standards for terminological redundancy or consistency, such terminological entries are not avoided or resolved as often as desirable. The only standards body we are aware of that pursues harmonization activities is IEC TC 1 \textit{Terminology} in cooperation with other TCs. However, specific harmonization actions may only be initiated by the TCs, not by TC 1.\footnote{In cooperation with other TCs, TC 1 tries to “integrate their terms and definitions that fall within the scope in [sic] the IEV, and in so doing to ensure to the greatest extent possible the correctness of the result” \cite{IECTC1.2017}. Correctness, in this case, is comprised by several factors, more precisely “that the IEV entries are consistent with each other, that each concept is identified by a single preferred term and different concepts are identified by distinct terms, and that the structural principles on which the IEV is based and other rules specified in the IEC Supplement to the ISO/IEC Directives, Annex SK, are respected” \cite{IECTC1.2017}. The objectives of TC 1 (stated in its strategic business plan, \citealp[cf.][]{IECTC1.2017}) clearly involve harmonization of IEV entries as well as the removal of doublettes. Until now, this has been applied to a subsection of the IEV (the 700s parts) and needs to be coordinated with each responsible TC.} Furthermore, the responsibility to avoid doublettes in the first place lies with single working groups that write the standards. The ISO/ IEC Directives, Part 2, 2016 \cite{ISO.2016}, for example, recommend the use of terminological databases which may already contain the concepts in question. The German Standardization Body DIN goes one step further with its standard DIN 820-2:2012 \cite{820-2.2012} by explicitly demanding that the repetition of concepts that have been defined elsewhere shall be avoided.\footnote{This standard is a modification of part 2 of the ISO/ IEC Directives from 2011. In the meantime, a new version has been published (DIN 820-2:2018-09, \citealp{820-2.2018}) which drops the explicit requirement not to repeat or define concepts anew. A reverse development can be expected from IEC: Current work on Annex SK in the IEC Supplement of the ISO/IEC Directives Part 1 is tending towards tightening the rule “one entry, one concept”, which also forbids a literal repetition of the same entry. Instead, a phrase referring to the original entry is supposed to be used.} What is more, the same concept should not receive multiple definitions in different standards. Third, variability in standardized terms and definitions may result from an omission of or a reluctance to use prior terminological systems and resources – especially when terminology work is considered to be of minor importance and conflicting priorities of standardization experts lead them to pursue other goals. A fourth reason may lie in the fact that terminology work requires a very close attention to detail as well as the ability to keep track of a rather complex system of many implicitly and explicitly interrelated items. Four hundred terminological entries – approximately the number of entries we allocated in our entry corpus for harmonization (see section \ref{sec-3.2}) – can be considered a huge amount of terminological items for a small group of experts. Terminology databases in standardization, however, may contain up to several thousand entries (e.g., the International Electrotechnical Vocabulary (IEV) of the International Electrotechnical Commission contains 20,000 terminological entries, the DIN-TERMinologieportal contains 750,000 entries, and an empty search in ISO’s Online Browsing Platform gives 216,205 results for all available languages (English, French, Russian, Spanish, and German). Of course, these huge sets naturally fall into smaller sets simply by belonging to different domains and sub-domains. However, a single technical committee, for example DIN NA 052 (\textit{Road Vehicle Engineering}) may still be responsible for almost 5,000 terms.\footnote{Please note that this is not the number of entries – since there is usually synonymy and homonymy in each domain-specific vocabulary, the number of terminological entries may be greater or smaller than the number of terms suggests. The exact number of terminological entries per technical committee, however, is not available via the portal.} Such numbers are a major obstacle for overall harmonization and quality assurance (see below). Fifth, some topics are cross-domain issues (e.g., safety and security) and concern a large number of standards bodies, which is another source for terminological variation. This last reason in particular is a motivator for our work, which is concerned with the harmonization of the concepts of IT security and functional safety for critical infrastructure. Standards bodies address this situation of convergence through institutional restructurings and alliances. As a result, a number of steering groups have been founded that interconnect hierarchically organized technical committees with each other and to external organizations.

In sum, a relevant sub-task of terminology standardization is the avoidance or resolution of doublettes. In practice, attempts to resolve doublettes are often term-based, that is, entries sharing the same term are identified and compared to see whether they are either redundant or necessary. Such a term-based procedure, unfortunately, misses all cases where different terms are used as identifiers of terminological entries. Instead, a thorough method would also have to take into account the fact that doublettes may be labeled by synonymous but different terms and would therefore have to search for similar definitions as well. Doublette management requires an assessment of linguistically variable terminological entries: Do several entries describe the same concept or do they not? Based on this decision, terminological entries could either be merged or need to be kept seperate. The notion of semantic similarity plays an important role in making such a decision: If two (or more) terminological entries are considered semantically similar enough, this may be a good indicator for conceptual identity, which justifies merging them or, in other words, to resolve doublettes. This is a much more time-consuming task than the term-based approach, which is why doublette resolution should be properly supported by tools.

A further requirement for terminologies is consistency, which can be stated as:

\begin{enumerate}[label={(\arabic*)}]
\setcounter{enumi}{6}
\item Similar and related terminological entries shall not unintentionally contradict any other terminological entry in form or content. \label{req7} 
\end{enumerate}

Because consistency within a corpus of terminological entries means avoiding all unwanted contradictions and maintaining a rigorous and unified theoretical and literal description, harmonization depends on the comparison of all terminological entries of one, usually two, or sometimes more \textbf{terminological resources} (= “text or data resource consisting of terminological entries” \cite{704.2009}). Because inconsistencies can arise in an unforeseen manner between any pair of entries within the terminology, consistency principally could only be guaranteed by comparing each entry with each other entry. Pairwise consistency $c$ between a pair of entries $(e_{i},e_{j})$, wherein $i,j ={1,… ,n}$, may be understood as an expert judgment over propositional and formal\footnote{Since it may cause confusion: when we use the adjective \textit{formal} we mean it along the lines of gloss 3.2 in Oxford Online Dictionaries “relating to linguistic or logical form as opposed to function or meaning” in a Saussurean sense or of gloss 1b in Merriam Webster “relating to or involving the outward form, structure, relationships, or arrangement of elements rather than content” (both accessed on 09.07.2018). In this sense, \textit{formal} is also applicable to the structure of terminological entries as described by TBX (cf. section 4.2). When applied to linguistic items traditionally described as “lexical”, we understand form in the simplest sense, that is, as a string of graphemes that make up a syntactical word (or parts thereof, cf. \citealp{Eisenberg.2013}) or a string of graphemes of a lexical word (i.e., an infinitive or nominative singular) or – in case of separable lexical words and syntactically complex lexemes – as all constituents belonging to the lexical expression (e.g., as in \textit{He \textbf{turned} the water \textbf{off}}.). \textit{Formal} can of course also relate to other characteristics of lexemes, for example gender or part of speech.} incoherence and coherence and shall be defined on a ratio scale within the range $-1 \leq c_{i,j} \leq 1$, wherein the zero point represents pairs of entries that differ thematically enough so that they would neither contribute to coherence nor incoherence. Overall consistency $C$ of the whole terminological resource shall then be defined as the sparse triangular matrix of all comparisons $c_{i,j}$ between all entries. Given self-consistency of each entry, this can be expressed as a unitriangular matrix:

\begin{gather}
\begin{bmatrix}
1 & & & &  0\\
c_{2,1} & 1 &&& \\
c_{3,1} & c_{2,1} & 1& &\\
\vdots & \vdots & &\ddots & \\
c_{n,1} & c_{n,2} & \cdots & c_{n,n-1} & 1
\end{bmatrix}
\end{gather}

For our corpus of terminological entries (or “entry corpus”, see section \ref{sec-3.2}) comprising only $n=446$ entries, the number of all theoretically necessary comparisons to guarantee consistency $C$, is thus

\begin{equation}
 n_{c}= \frac{n(n-1)}{2}= 99,235
\end{equation}

Guaranteeing pairwise consistency would therefore be an overwhelming endeavor that heavily depends on time investments of the participants and a meticulous overview of all (relevant) entries, while contemplating the referred concepts and thus scientific, factual, and regulative concerns. Even more so, considering that presumably most of these comparisons are fairly meaningless in and of themselves and therefore irrelevant for a consistency task. Though ad-hoc heuristics depending on the inherent structure of the domain, as well as methods depending on the conceptual structure of the terminology under harmonization, may help to identify relevant entries, it is very hard to predict important comparisons and safely ignored comparisons in an unbiased manner.\footnote{A method for terminology harmonization depending on conceptual systems is specified by ISO 860:2007 \cite[cf.][]{860.2007}, a standard that is applicable to either the national or the international level and can therefore be used both in monolingual and multilingual contexts. This standard applies an approach of conceptual comparison that is guided by the heuristic of the onomasiological structure, that is, the concept system. Here, it is especially important to identify the onomasiological structures of different terminological resources, that is, the relations between the concepts and – in case of hierarchical relations – criteria for subdivision. Only after these factors of divergence in terminological resources have been identified, harmonization proceeds to the level of single concepts and only after that to the level of terms and definitions. Unfortunately, applying this method of harmonization to terminology in standardization is mostly bound to single standards and requires an effort in reconstruction. Standardized terminologies are scattered over documents so that there is no such thing as a huge terminological system with explicitly specified relations between the entries. At best, relations played a role while working on a standard’s terminology so that it may show more or less explicit in the definitions and the systematic order of the standard. Exceptionally, some standards contain an overview of the relations between the concepts (visually displaying some unlabeled edges between labeled nodes, sometimes using UML notation to indicate the relation type). Harmonization would therefore start with the reconstruction of the concept system, optionally its visualization and would then properly start by comparing this onomasiological macrostructure of one terminological resource to another one with potential overlap. Finding all direct semantic relations in the haystack of candidate pairs would ultimately require the same traversal of the unitriangular matrix, while determining not the degree of consistency but the type of semantic relation. Wherever a large corpus of concepts has to be considered, the relational nature of concept systems causes quadratic increase of work in relation to corpus sizes.}

Thus, it is paramount to overcome quadratic increases of complexity in larger corpora of terminology. In light of recent accomplishments in computational linguistics, appropriate support for consistency management as well as for doublette resolution should be provided by software helping to capture the relatedness of a set of terminological entries, or rather of the concepts they describe. This software should operate on semantic representations of terminological data, since a concept may have several preferred designations in different terminological entries so that semantic information retrieval is mandatory. This, however, requires an evaluation of semantic representations for terminological data, which are by definition domain-specific. We currently address the issues introduced so far for the terminology of a specific domain, namely that of electrotechnical standards for functional safety, IT security and dependability of critical infrastructure. Standardization, however, is not limited to a single domain so that a distributional semantic model to support harmonization tasks needs to be applicable to other domains as well. With this requirement in mind, we apply the model of \citet{Arora.2017} to support our tasks, which is introduced in section \ref{sec-4}. We chose it to appropriately face the peculiarities of terminological entries, which are described in section \ref{sec-4.2}. To evaluate the reliability of this model, we chose a similarity task. This similarity task employs terminological entries from our domain. Before we introduce our evaluation data set and the methods applied for its construction in section \ref{sec-3}, we will discuss related work in section \ref{sec-2}. In section \ref{sec-5} we will discuss the results of the evaluation of the embedding model and will come back to the evaluation data set in section \ref{sec-6} where we review its strengths and shortcomings and join a more general debate about intrinsic evaluation methods. We will conclude with an outlook for future work.

\section{Related work}
\label{sec-2}

Embeddings, semantic representations for linguistic entities ranging from lexical items to texts, have become increasingly popular for general-purpose, that is, non-domain-specific applications. Methods like Latent Semantic Analysis \cite{Dumais.2004} have exploited the co-occurrence of words in a corpus to represent the meaning of a word as a vector, wherein the number of co-occurrences with each other word in the vocabulary is counted. Recent approaches use neural networks to create word embeddings, condensing the number of dimensions per word from the size of the vocabulary to a few hundreds. From the very popular approach word2vec \cite[cf.][]{Mikolov.2013} a whole family of 2vec-approaches has evolved (\citealp[e.g., doc2vec by][]{Le.2014}, \citealp[phrase2vec by][]{Cho.2014}, and many more) as well as a great number of other models that are not 2vec-based \cite[e.g.,][]{Pennington.2014, Wieting.2016,Arora.2017}. In addition, there is high interest in using embeddings for domain-specific applications as well (\citealp[e.g., in IT security, cf.][]{Roy.2017,Han.2017}, \citealp[in medicine, cf.][]{Banerjee.2017, Ghosh.2016}, but also in other domains). Quite a number of domain-specific approaches try to capture the semantics of domain-specific vocabulary (i.e., terms) by training vector representations on domain-specific corpora since general-purpose corpora may not address topics of highly specialized domains in very much detail. \citet{Sugathadasa.2017} report that domain-specific embeddings outperform general-purpose embeddings. A general problem with domain-specific corpora, however, is that they may not meet requirements for token size. In a study of \citet{Robin.}, this lead to the rejection of embeddings trained on domain-specific corpora in favor of embeddings trained on general-purpose corpora, since these performed better in a down-stream task. Furthermore, \citet{Roy.2017} claim that domain-specific corpora may not suffice to guarantee reliable domain-specific embeddings, since even their items can be sparse and not (co-)occur often enough although they are of relevance to the domain.

To account for the sparseness of terms in general-purpose and domain-specific corpora, more structured forms of domain-specific resources are often introduced into the training of embedding models to optimize the vectors (e.g., \citealt{Roy.2017, Wang.2015, Ghosh.2016})\footnote{For example, the Unified Medical Language System (UMLS, \href{https://www.nlm.nih.gov/research/umls/}{https://www.nlm.nih.gov/research/umls/}), the Common Vulnerabilities and Exposures data set in the National Vulnerability Database provided by NIST (\href{https://nvd.nist.gov/}{https://nvd.nist.gov/}) or even “handmade” vocabularies consisting of lists of terms.\label{footnote10}}. This, of course, implies a step from unsupervised learning methods to semi-supervised or supervised learning methods. Alternatively, semantic representations other than vector-based ones can be used \cite[e.g.,][]{Pedersen.2007}. A third option would be weighting approaches as will be applied here (cf. section \ref{sec-4}).

Interestingly, in quite a number of studies embeddings for terms in the domain of IT security are created, see for example \citet{Han.2017}, \citet{Roy.2017}, or \citet{Mittal.}. These are unsupervised models trained on IT-security-related documents. Optionally, these models integrate domain-specific knowledge as input to the model training.\footnote{See footnote \ref{footnote10}; futhermore knowledge graphs like the Cybersecurity Information Ontology by \citet{Takahashi.}, or the Unified Cybersecurity Ontology (UCO) by \citet{Syed.2016} are used.} For example, \citet{Roy.2017} annotate vulnerability and malware terms with information from graph-based knowledge resources and subsequently train their model with this type of input. For our purposes, these models are not sufficient, since they favor specific types of IT-security-related vocabulary, while the terminology in standardization is concerned with very basic principles of IT security and operates on a higher onomasiological level.

A further problem for domain-specific vector representations is the evaluation of their reliability. Several evaluation data sets are available that comprise (lexical) items from general, everyday language (\citealp[e.g., MEN by][]{Bruni.2012}, \citealp[SimLex-999 by][]{Hill.2015}, \citealp[and WordSim-353 by][]{Finkelstein.2001}). For our domain-specific purposes, these data sets are not sufficient for evaluation, so that we created our own data set for similarity evaluation, Harbsafe-162. It includes textual information in the shape of definitions, thereby leveraging distinctive information that is prominently available in standardization. A common method to create evaluation data sets is to conduct a rating task on pairs of linguistic items. This method of intrinsic evaluation has found entry into quite a number of SemEval-tasks, for example \citet{Agirre.2012}, \citet{CamachoCollados.2017}, \citet{Cer.2017}, or \citet{Jurgens.2014}. The rater is supposed to compare both items and make a judgement on their semantic similarity. This similarity judgment can in turn be correlated with the distance of the embeddings or, more commonly, with their cosine similarity. If the similarity given by human raters and the distance between computed vector representations fits, the vector representations are considered reliable. The rating task usually requires the participation of several raters so that high inter-annotator agreement can be proven and arbitrariness and randomness of the ratings can be ruled out. In domain-specific contexts, the rating task should not be performed by out-of-domain speakers (as for example found at Amazaon’s Mechanical Turk). Raters should rather be domain experts. Those, however, are usually not easily available for a time-consuming task. Some studies therefore strive for other evaluation techniques. Since we perform our work in a collaborative project that is connected to domain experts, we will nevertheless use this method to evaluate embeddings for terminological entries from different standards. We give a detailed description of our procedure in section 3. Since to our knowledge there is no other work on embeddings for terminological entries, we freely adapted the method as described by \citet{CamachoCollados.2017} to our specific case. It is itself an adaption of \citet{Jurgens.2014}, where linguistic items on different levels (e.g., lexical items, paraphrases, sentences and so on) were compared.

\section{Evaluation data set design - Harbsafe-162}
\label{sec-3}
With our domain-specific focus, we try to develop vector representations that can serve the purpose of terminological entry harmonization as described in section 1. Our data therefore originate from specific sources, namely electrotechnical standards that deal with issues of functional safety, IT security, and dependability of critical infrastructure. The corpus of all relevant standards consists of approximately 805 texts. For the creation of our data set for embedding evaluation, we chose only nine texts \cite{62443-1-1.2009, 62443-3-1.2009, 61508-4.2010, 62443-2-1.2010, 62443-3-3.2013, Guide.51, 62443-2-3.2015, 62443-2-4.2015, Guide-120-D.2018} since they are especially relevant for the topic area. We call this collection of standards our “standards corpus”.

\subsection{Rating scale}
\label{sec-3.1}
As mentioned above, our evaluation data set is based on a rating task in which raters were asked to assess the similarity of two terminological entries at a time. We used a rating scale based on the scale in SemEval 2017 Task 2 \cite[cf.][]{CamachoCollados.2017} for this is a scale already adapted to lexical data (cf. Table \ref{table1}).

\begin{table}[h]
\caption{Five-point Likert scale used for word similarity ratings in SemEval 2017 Task 2 \cite{CamachoCollados.2017}}
\resizebox{0.75\textwidth}{!}{
\begin{minipage}{\textwidth}
\renewcommand{\arraystretch}{1.5}
\begin{tabular}{p{0.5cm}p{3cm}p{13cm}}
\hline
4 & Very similar & The two words are synonyms (e.g., \textit{midday}-\textit{noon} or \textit{motherboard}-\textit{mainboard}).\\ 
\hline 
3 & Similar & The two words share many of the important ideas of their meaning but include slightly different details. They refer to similar but not identical concepts (e.g., \textit{lion}-\textit{zebra} or \textit{firefighter}-\textit{policeman}). \\ 
\hline 
2 & Slightly similar & The two words do not have a very similar meaning, but share a common topic/domain/function and ideas or concepts that are related (e.g., \textit{house}-\textit{window} or \textit{airplane}-\textit{pilot}). \\ 
\hline 
1 & Dissimilar & The two words describe clearly dissimilar concepts, but may share some small details, a far relationship or a domain in common and might be likely to be found together in a longer document on the same topic (e.g., \textit{software}-\textit{keyboard} or \textit{driver}-\textit{suspension}). \\ 
\hline 
0 & Totally dissimilar and unrelated & The two words do not mean the same thing and are not on the same topic (e.g., \textit{pencil}-\textit{frog} or \textit{PlayStation}-\textit{monarchy}). \\ 
\hline 
\end{tabular}
\end{minipage} }
\label{table1}
\end{table}

This scale had to be adapted to our specific purposes for four reasons (cf. Table \ref{table2}). First, we use specific kinds of data, namely terminological entries. Terminological entries are comprised of one term, several synonymous terms or term phrases and a definition. The terms can range from monomorphematic items to compositional or phrasal multi-word expressions (e.g., ADJ + NN). The definition usually is a syntactic structure that may replace the term in a text and explains the sense of the term(s). The scale of \citet{CamachoCollados.2017} on the other hand, was tailored to their data, that is, pairs of words without glosses explaining the senses of these words. Second, since our data consist of terminological entries, it is not in the strictest sense correct to call a relation between two entries synonymy. Still, it comes very close: Since terminological entries represent items of a domain-specific concept system and there may be several terminological entries representing one concept slightly differently, it comes very close to the relation between linguistic lexical signs. The relation between two rating items in our data set, however, may be more adequately described in terms of conceptual identity, conceptual similarity, and relatedness. The scale is therefore terminologically adapted to our evaluation task (cf. category 4 in Table \ref{table1} and \ref{table2}). Third, these modifications were also motivated by the fact that we did not want participants to focus on isolated words – in our case terms – only and disregard the definition, but to consider both. Instead of using the expression \textit{terminological entry}, though, we used the more comprehensible term \textit{concept}, since that is what terminological entries are supposed to represent (see section \ref{sec-1}). Fourth, this, in turn, also supported our attempt to provide raters with information about the goal of the rating task.\footnote{You can find the online rating tool at \href{https://harbsafe.tu-braunschweig.de}{https://harbsafe.tu-braunschweig.de}.}

\begin{table}[h]
\caption{Five-point Likert scale for terminological entry similarity ratings (for both English and German participants)}
\resizebox{0.75\textwidth}{!}{\begin{minipage}{\textwidth}
\renewcommand{\arraystretch}{1.5}
\begin{tabular}{p{0.5cm}p{2,8cm}p{5cm}p{2.8cm}p{5cm}}
\hline
4 & Very similar & Both concepts are semantically very similar or identical (e.g., \textit{midday}-\textit{noon} or \textit{motherboard}-\textit{mainboard}). & Sehr ähnlich & Die beiden Begriffe sind inhaltlich sehr ähnlich oder identisch (z.B. \textit{midday} – \textit{noon} oder \textit{motherboard} – \textit{mainboard}).\\ 
\hline 
3 & Similar & Both concepts have many semantic similarities but are different in details. They are closely related to each other (e.g., \textit{lion}-\textit{zebra} or \textit{firefighter}-\textit{policeman}).  & Ähnlich & Die beiden Begriffe haben viele inhaltliche Ähnlichkeiten unterscheiden sich aber in Details. Sie stehen in einem sehr engen Zusammenhang (z.B. \textit{lion} – \textit{zebra} oder \textit{firefighter} – \textit{policeman}).\\ 
\hline 
2 & Slightly similar & Both concepts are semantically not very similar but belong to the same topic or domain. They are directly related to each other (e.g., \textit{house}-\textit{window} or \textit{airplane}-\textit{pilot}). & Etwas ähnlich & Die beiden Begriffe sind inhaltlich nicht sehr ähnlich, gehören aber zum selben Thema oder Sachgebiet. Sie stehen in einem direkten Zusammenhang (z.B. \textit{house} – \textit{window} oder \textit{airplane} – \textit{pilot}). \\ 
\hline 
1 & Dissimilar & Both concepts are semantically not similar but belong to the same topic or domain. They are indirectly related to each other (e.g., \textit{software}-\textit{keyboard} or \textit{driver}-\textit{suspension}).  & Unähnlich & Die beiden Begriffe sind inhaltlich nicht ähnlich, gehören aber zum selben Thema oder Sachgebiet. Sie stehen in einem nur entfernten Zusammenhang (z.B. \textit{software} – \textit{keyboard} oder \textit{driver} – \textit{suspension}).\\ 
\hline 
0 & Totally dissimilar and unrelated & Both concepts are semantically not similar and do not belong to the same topic or domain. They are not related to each other (e.g., \textit{pencil}-\textit{frog} or \textit{PlayStation}-\textit{monarchy}). & Vollkommen unähnlich und nicht zusammenhängend & Die beiden Begriffe sind inhaltlich nicht ähnlich und gehören nicht zum selben Thema oder Sachgebiet. Sie stehen in keinem Zusammenhang (z.B. \textit{pencil} – \textit{frog} oder \textit{PlayStation} – \textit{monarchy}).\\
\hline
\end{tabular}
\end{minipage} }
\label{table2}
\end{table}

In our rating task, we did not allow for intervals on the scale (e.g., .5 or .25 steps) since we consider the relations as distinct, ordered categories, not as intervals. We understand the rating scale as ordinal in this article.

\subsection{Selection of terminological entry pairs}
\label{sec-3.2}

\begin{figure}[h]
\centering
\label{fig.1}
\begin{subfigure}{0.49\textwidth}
\centering
\includegraphics[scale=0.99]{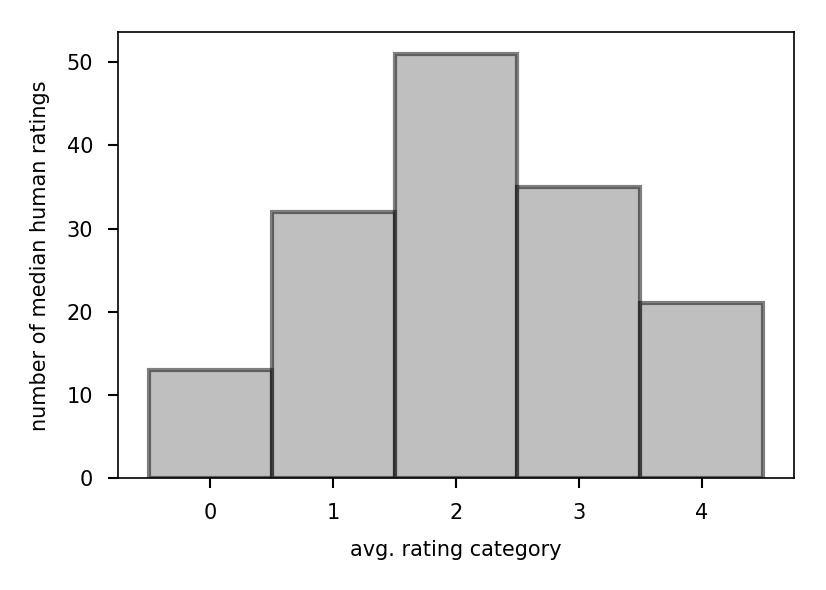}
\captionsetup{width=0.85\linewidth}
\caption{Distribution of \emph{averaged} ratings for all entry pairs (both raters).}
\label{fig.1a}
\end{subfigure}
\begin{subfigure}{0.49\textwidth}
\centering
\includegraphics[scale=0.99]{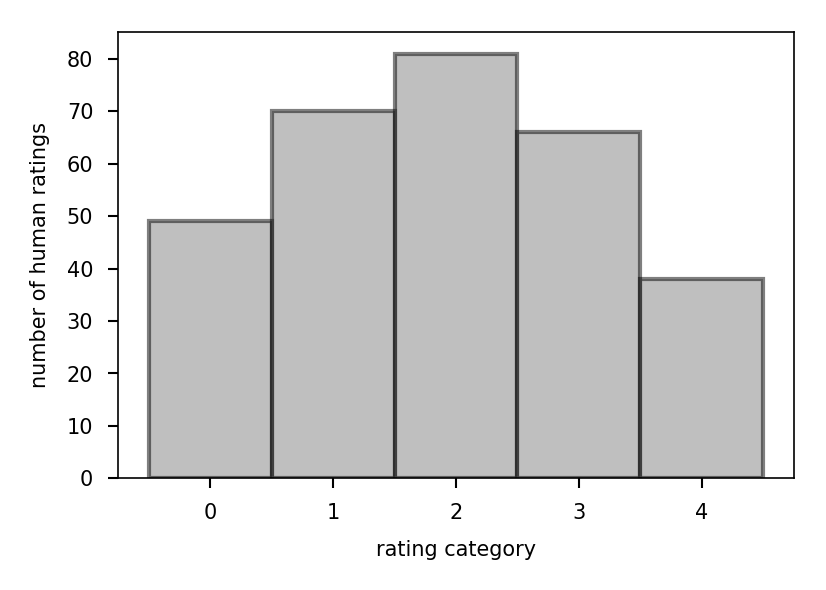}
\captionsetup{width=.85\linewidth}
\caption{Distribution of \emph{non-averaged} ratings for all entry pairs (both raters).}
\label{fig.1b}
\end{subfigure}
\caption{Rating distributions.}
\end{figure}

From our standards corpus, we took all terminological entries contained in these standards as candidate entries – overall, this “entry corpus” contains 446 terminological entries. From this entry corpus, we created the actual sample for Harbsafe-162 with three requirements in mind. First, pairs should be distributed more or less evenly across the similarity scale or rather no single category should be dominant over all other categories (to avoid prevalence, cf. \citealt{Artstein.2008}). Second, the sample should comprise a sufficiently large number of pairs. Third, the sample should comprise only a number of pairs that is manageable by volunteer domain experts. The latter two led us to fix a data set of 152 pairs with 258 out of the 446 terminological entries from our entry corpus. Harbsafe-162 is a comparatively small evaluation data set. The selection of these 152 pairs was conducted as an integrated selection and rating process, in which we performed the first ratings ourselves, using the same scale as our volunteer experts in the actual rating procedure. Independent of each other, each of us chose pairs from the entry corpus and rated them. This was guided by a provisional implementation of the similarity rankings detailed in section \ref{sec-4.2}, which allowed us to pick one entry and then choose other entries  from the ranking provided by the provisional implementation. The single entries in this ranking were tagged for a topic, each of which had a distinct color. The process of pair selection was continually accompanied by a chart summarizing the number of pairs that have been rated with a specific category of our rating scale. There was no information on how single pairs have been rated, however. After we had reached the number of 152 pairs, each of us rated the pairs suggested by the other author. Certain entries were not considered for the data set (e.g., entries that contain references to external sources, figures that would not be available to other raters or entries that do not contain a definition but only a synonym). Entries containing acronyms were slightly modified by provision of the full form. Figure \ref{fig.1a} shows the distribution of pairs among averaged categories; Figure \ref{fig.1b} shows the distribution of individual ratings by both raters.

Considering the fact that each category was represented at least ten times and pairs did not fall into one single category (cf. Figure \ref{fig.1a}), we saw our first requirement as fulfilled. Our inter-annotator agreement on similarity ratings was 0.78 measured by Krippendorff’s $\alpha$ \cite{Krippendorff.}. Variation of values assigned to each pair keeps within limits: A difference of three categories was observed for two pairs (1\%), a difference of two categories for only six pairs (4\%), a difference of one category for 62 pairs (41\%) and no difference was observed for 82 pairs (54\%). A variation of four categories was not observed (cf. Table \ref{table3}).

\begin{table}[h]
\caption{Contingency table for user 12 and user 13 (i.e. for the author ratings)}
\centering
\resizebox{0.75\textwidth}{!} {\begin{minipage}{\textwidth}
\renewcommand{\arraystretch}{1.5}
\begin{tabular}{c p{1.5cm} p{1.5cm} p{1.5cm} p{1.5cm} p{1.5cm} p{1.5cm} p{1.5cm}} \cline{2-6}
& \multicolumn{5}{c}{user 13}\\
\cline{1-6}
\multicolumn{1}{c|}{user 12} & \multicolumn{1}{c}{4} & 3 & 2 & 1 & 0 & \multicolumn{1}{c}{}\\
\hline
\multicolumn{1}{c|}{4} & \textbf{16} & 5 & 0 & 1 & 0 & \multicolumn{1}{c}{22}\\
\hline
\multicolumn{1}{c|}{3} & 0 & \textbf{22} & 6 & 3 & 1 & \multicolumn{1}{c}{32}\\
\hline
\multicolumn{1}{c|}{2} & 0 & 6 & \textbf{21} & 15 & 2 & \multicolumn{1}{c}{44}\\
\hline
\multicolumn{1}{c|}{1} & 0 & 1 & 10 & \textbf{10} & 13 & \multicolumn{1}{c}{34}\\
\hline
\multicolumn{1}{c|}{0} & 0 & 0 & 0 & 7 & \textbf{13} & \multicolumn{1}{c}{20}\\
\hline
\multicolumn{1}{c}{} & 16 & 34 & 37 & 36 & 29 & \multicolumn{1}{c}{\textbf{152}}\\ \cline{2-7}
\end{tabular}
\end{minipage} }
\label{table3}
\end{table}

We therefore decided to take the next step with Harbsafe-162 by involving project-external raters recruited from academia, industry, and standardization.

\subsection{Rater recruitement and procedure}
\label{sec-3.3}
We chose to recruit volunteer domain experts as raters through several channels. On the one hand, we tried to address experts by publications in standardization-related media \cite{Arndt.2018}. Furthermore, we made the request public on our institutional Web sites.\footnote{(1) \href{www.iva.ing.tu-bs.de}{www.iva.ing.tu-bs.de}, (2) \href{http://www.iva.ing.tu-bs.de/?iT=4\&projectId=557}{http://www.iva.ing.tu-bs.de/?iT=4\&projectId=557}, (3)~\href{https://www.dke.de/de/themen/projekte/harbsafe}{https://www.dke.de/de/themen/projekte/harbsafe}} On the other hand, we made use of expert networks. Via the system safety mailing list, selected domain experts have been invited to participate.\footnote{\href{http://www.systemsafetylist.org/}{http://www.systemsafetylist.org/}} Additionally, we invited members of the VDE YoungNet to participate.\footnote{\href{https://www.vde.com/de/vde-youngnet}{https://www.vde.com/de/vde-youngnet}} Members of standardization organizations have been informed via presentations or their respective distribution lists. Participation was anonymous but required registration on our online rating tool and the use of one of multiple codes that indicate the recruitment method.

\subsubsection{Rating task instructions}
\label{sec-3.3.1}

\begin{figure}[h]
\centering
\frame{\includegraphics[scale=0.5]{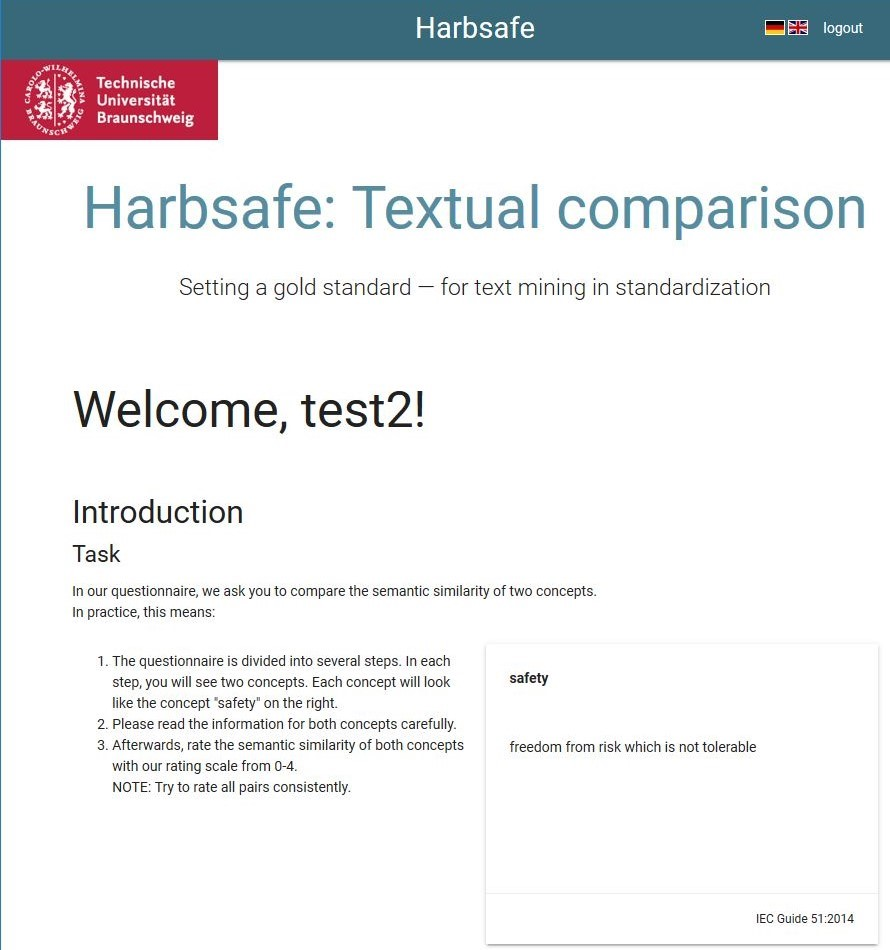}}
\caption{Harbsafe evaluation tool – screenshot 1}
\label{fig.2}
\end{figure}

Raters were given a short introduction to the task, consisting of three parts. First, we described the task itself in three steps (cf. Figure \ref{fig.2}). Initially, we explained what kind of information we would present to the raters: two concepts, each represented by one or several terms and a definition. Therefore, our data supply the raters with context that disambiguates the terms explicitly (instead of implicitly, as would be the case with random co-text). Here, we already showed an example of such a concept. Second, we asked our raters to read the information on both concepts carefully. Third, we told them what we expected them to do after reading: “Rate the semantic similarity of both concepts with our rating scale from 0-4. NOTE: Try to rate all pairs consistently.” This instruction has been provided in German or English.

\begin{figure}[h]
\centering
\frame{
\includegraphics[scale=0.6]{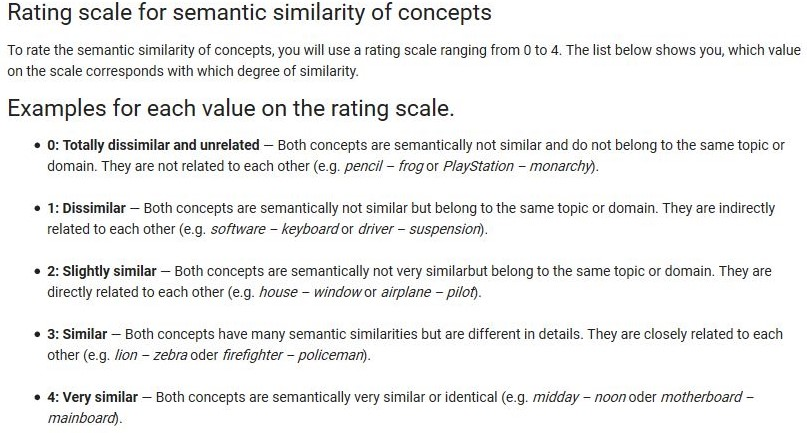}}
\caption{Harbsafe evaluation tool – screenshot 2}
\label{fig.3}
\end{figure}

After the task description, we introduced our rating scale as shown in Table 2 to familiarize the raters with it. Two example word pairs, which can be characterized as parts of everyday language, accompanied the explanation of each scale point. This was to ensure that raters could easily relate the definition of the scale point to the example and use the scale as intended (cf. Figure \ref{fig.3}).

\begin{figure}[h]
\centering
\frame{
\includegraphics[scale=0.7]{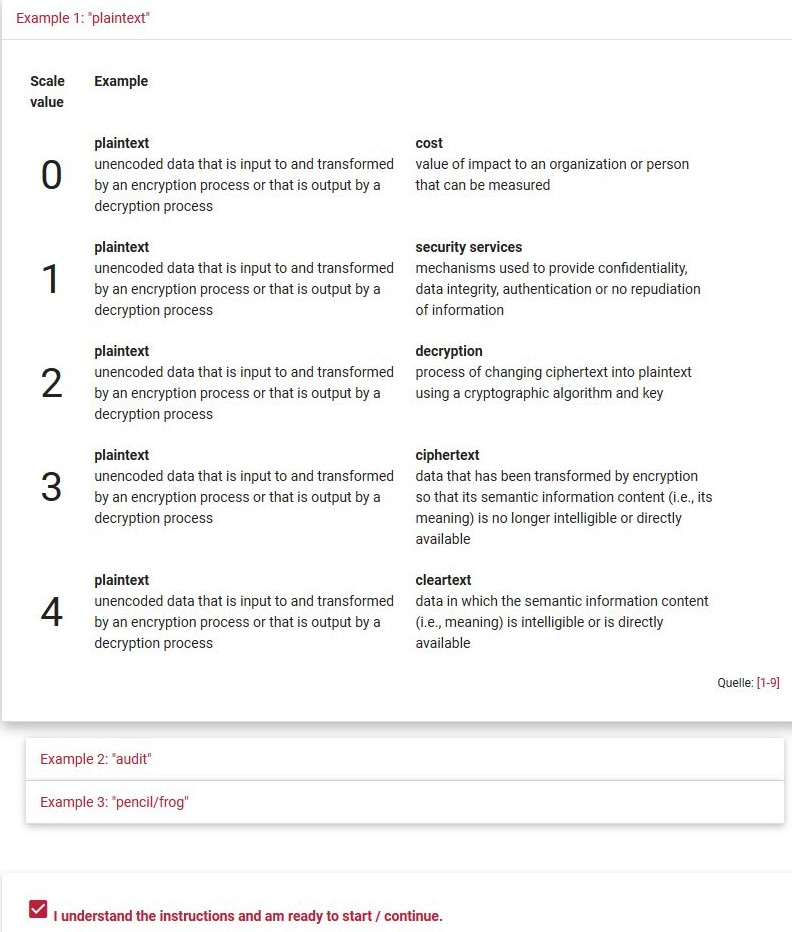}}
\caption{Harbsafe evaluation tool – screenshot 3}
\label{fig.4}
\end{figure}

Since the examples in the explanation of our scale were only word pairs, not terminological entries, we presented these examples again in the form of the data in the actual rating task, that is, as words/ terms and glosses/ definitions. Since our data are very abstract domain-specific concepts, we added two additional sets of examples from the entry corpus that are not included in Harbsafe-162. This way, raters would already be confronted with very abstract and complicated concepts before having to use the scale themselves. We tried to keep variation as low as possible in these pairs to demonstrate the change of relation between them and nothing else. The first item of these domain-specific pairs was therefore identical over all scale points. Furthermore, we hoped that this multiplication of examples would provide every rater with examples he could relate to thereby fostering learning by example and avoiding inappropriate use of the scale. The additional examples were optionally presented, thus raters could consult them at will. Last, we asked raters to confirm that they had understood the instructions (cf. Figure \ref{fig.4}).

Raters had the option to finish the rating process over several sessions and to postpone any decision. They were not allowed, however, to modify their ratings in retrospect.

\subsubsection{Rating process of volunteer domain experts}
\label{sec-3.3.2}

Raters have been provided with Harbsafe-162 in random order without clusters of same-similarity pairs. Additionally, approximately every sixteenth comparison item was from everyday language (on top of the 152 pairs of Harbsafe-162). These pairs have been taken from the data set in \citet{CamachoCollados.2017}. Since this data set only contains word pairs, glosses for each word were added from Oxford Online Dictionaries or WordNet to make these examples comparable to the terminological pairs. These additional pairs served two purposes: First, compared to the terminological pairs, they are easily understood and can thus serve as an interruption in our highly demanding rating task and provide raters with motivation. Second, these “easy-as-pie-pairs” are an additional means of quality control and can indicate problems with our methodology (cf. section \ref{sec-3.3.3}). Since we chose these pairs as prototypes for each rating scale category, they have an intended rating that should be given by our raters. Deviations, however, indicate that there may be problems with our rating scale, that is, raters may not have understood it. They may also indicate problems with our data set: If ratings among the easy pairs are highly reliable as expected, but Harbsafe-162 items are unreliable, we can probably exclude that raters have not understood the rating scale.

\subsubsection{Rating results of all raters (including the authors)}
\label{sec-3.3.3}

Our efforts to recruit domain experts as raters were comparatively successful: Overall, four domain experts completed the rating task so that we come to six raters – us included. Judging from the logs, the recruited raters needed on average approximately 43 minutes to complete the task, which amounts to nearly 4 pairs per minute. We were surprised about the impressive display of spontaneity, which may be explained by the familiarity of the experts with the underlying concepts. It casts doubt though, whether they felt a need to read the definitions carefully. To assess the performance of each rater we computed (1) Krippendorff’s $\alpha$ \cite[cf.][]{Krippendorff.} of their ratings to all the others’ median ratings and (2) the average pairwise Spearman correlations between their ratings and those of the others’ ratings (cf. Table \ref{table4}). Since user 23 falls outside one standard deviation of the mean in both measures, we considered to remove their data from the rating set. Before doing so, however, we tried to confirm this decision by referring to other performance criteria. First, we counted each agreement of a rater with any other rater above a threshold level of 0.7 (by Krippendorff’s $\alpha$). This gave the performance ranking in Table \ref{table5}. Again, user 23 is on a low performance rank as compared to all other raters.

\begin{table}[h]
\caption{Rater performance}
\centering
\begin{subtable}[h]{.5\textwidth}
\captionsetup{width=.9\linewidth}
\caption{Rater performance by agreement with median ratings (Krippendorff’s $  \alpha$) and average pairwise Spearman correlations to all other raters.}
\resizebox{0.75\textwidth}{!} {\begin{minipage}{\textwidth}
\renewcommand{\arraystretch}{1.5}
\begin{tabular}{ccc}
\hline
\multicolumn{1}{p{1cm}}{Rater} & \multicolumn{1}{p{3cm}}{Krippendorff's $ \alpha $} & \multicolumn{1}{p{3cm}}{Average pairwise Spearman's $\rho$} \\
\hline
user 22 & 0.81 & 0.75\\ \hline
user 12 & 0.80 & 0.74 \\ \hline
user 13 & 0.76 & 0.68  \\ \hline
user 24 & 0.74 & 0.73 \\ \hline
user 26 & 0.71 & 0.73 \\ \hline
user 23 & 0.66 & 0.63 \\ \hline
\end{tabular}
\end{minipage} }
\label{table4}
\end{subtable}%
\begin{subtable}[h]{.5\textwidth}
\captionsetup{width=.9\linewidth}
\caption{Rater performance by agreement with median ratings (Krippendorff’s $  \alpha$) and average pairwise Spearman correlations to all other raters.}
\resizebox{0.75\textwidth}{!} {\begin{minipage}{\textwidth}
\renewcommand{\arraystretch}{1.5}
\begin{tabular}{ccc}
\hline
\multicolumn{1}{p{1cm}}{Rank} & \multicolumn {1}{p{3cm}}{Rater} & \multicolumn{1}{p{3cm}}{Number of agreements $>0.7$}\\
\hline
1 & user 12 & 3\\ \hline
1 & user 22 & 3 \\ \hline
2 & user 13 & 2  \\ \hline
3 & user 24 & 1 \\ \hline
3 & user 26 & 1 \\ \hline
4 & user 23 & 0 \\ \hline
\end{tabular}
\end{minipage} }
\label{table5}
\end{subtable}
\end{table}

As a final measure to assess the understanding of our rating scale, we used our easy-as-pie-pairs for which we expect very good performance. If raters fail them, they may not have used the rating scale as intended and therefore their data might be unreliable. An important question for this performance measure needs to be answered, though: Which is worse? Frequently using the scale in an unintended manner but with low deviations, or having few but huge deviations from its intended use? Even though we achieved 100\% agreement in our own ratings for the easy-as-pie-pairs, we decided that infrequent but larger deviations make the difference. As explained above, small deviations were fairly common in our own ratings of the Harbsafe-162 pairs: Deviations of one point were observable for 41\% of all pairs. The number drops considerably for deviations of two points, which was observable for 4\% of all pairs and gets even lower for higher deviations. We therefore decided to judge the raters’ performance by the number of deviations of two or more scale points from the rating the easy-as-pie-pairs were supposed to have, had the scale been applied in the intended way. Here, again, user 23 has the lowest rank because they have one deviation of two or more scale points in the easy-as-pie-pairs while all other raters have none. We therefore excluded user 23’s ratings from Harbsafe-162.

After removal of user 23, overall inter-rater agreement measured by Krippendorff’s $\alpha$ thus increases from $\alpha=0.67$ to $\alpha=0.7$. Average pairwise Spearman correlation between the ratings of all respondents increases from $\rho=0.72$ to $\rho=0.74$. These figures are quite high: Lexical data sets report agreements of $\rho=0.67$ in the case of SimLex-999 \cite{Hill.2015} or $\rho=0.61$ for WS-353 \cite{Finkelstein.2001}. However, these only comprise words, while Harbsafe-162 also includes synonyms and definitions that disambiguate the meaning and give context to any rating decision. Thus, a higher agreement should be expected for Harbsafe-162 versus other lexical evaluation sets. On the other hand, rating tasks on semantic textual similarity of extracted headlines, Q\&As, beliefs or image captions tend to achieve comparable or higher agreements \cite[cf.][]{Agirre.2015}. However, differences in the rating scale and the likely easier task to identify propositional congruence may render the values incomparable. Overall, we regard the achieved inter-rater agreement as a good indication for adequate reliability.

\begin{figure}[hbt]
\centering
\includegraphics[scale=1]{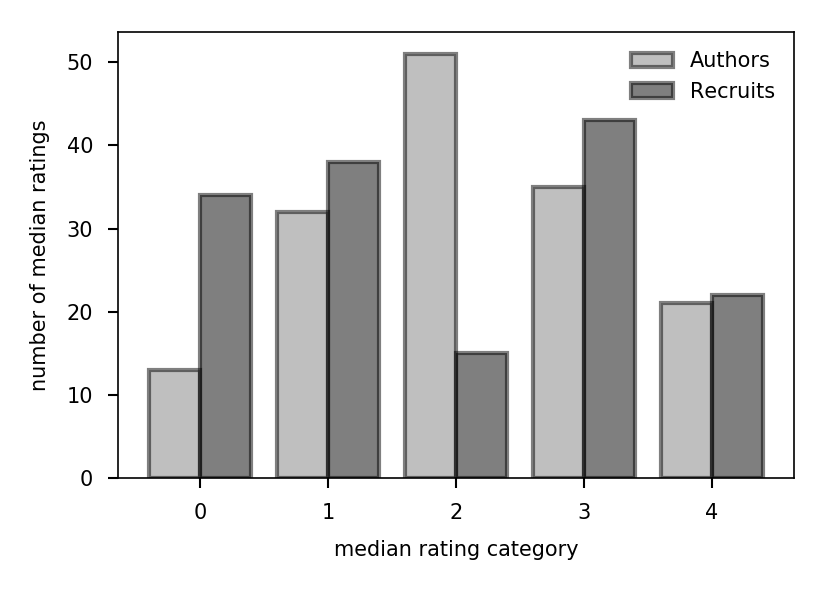}
\caption{Comparison of number of median author ratings and median domain expert ratings (0.5 ratings are rounded up).}
\label{fig.5}
\end{figure}

Figure \ref{fig.5} compares the number of median author ratings and median domain expert ratings. There is a noticeable mismatch for categories 0 and 2 when comparing the author ratings to the median ratings of the domain experts. When the domain experts tended to state that there is no relation whatsoever between entries (33 pairs), the authors often understood both concepts to be indirectly related (17 pairs, e.g., \textit{untrusted channel} – \textit{harmful event}, \textit{diversity} – \textit{use case}). Only in few cases did one of the authors see a direct relation (5 pairs, e.g., \textit{security incident} – \textit{enterprise}). As regards the number of author ratings in category 2, the domain experts showed a wider spectrum: Two times they rated for 4, eleven times they rated for 3, four times they rated for 2, twenty-three times they rated for 1 and six times they rated for 0 when the authors chose category 2. We will touch on possible reasons for this in section \ref{sec-6} where we discuss probable limitations of Harbsafe-162 and join the discussion on intrinsic evaluation data sets.

\section{Semantic representations for terminological entries and their evaluation}
\label{sec-4}

After having presented Harbsafe-162, we will now demonstrate its application in an evaluation of a semantic model of our entry corpus. Primarily, we try to estimate the predictive power of the model for future use cases within the Harbsafe project. In addition, however, we will also gather information for further optimizing Harbsafe-162 in the future. In section \ref{sec-4.1} we will provide a more detailed characterization of the data we try to create representations for. The nature of terminological entries motivated us to use the model by \citet{Arora.2017}, which will be described in section \ref{sec-4.2}. In sections \ref{sec-4.3} and \ref{sec-4.4} we will then report on the evaluation experiment.

\subsection{Characteristics of terminological data}
\label{sec-4.1}

Terminological data show several characteristics that need to be considered when creating their semantic representations with distributional semantic models as well as assessing their reliability. Some of these characteristics are of course also relevant for lexical data.

Characteristic 1. Terminological entries are by definition domain-specific. This is a crucial characteristic, since domain-specific items might be underrepresented in freely available general corpora and lexical resources on which distributional and other semantic models are trained.

Characteristic 2. Terminological entries go well beyond the purely lexical information that is encoded by a term or a set of synonymous or equivalent terms. Above, the definition was already mentioned as a relevant piece of information that may be used while training a distributional semantic model. An entry, however, is a hierarchy of structured data as specified by the xml-based format TBX \cite[cf.][]{30042.2008}. At most, all of the data of an entry may be part of its semantic representation. A distributional semantic model needs to make a unified vector representation of the bag of words taken from the TBX-entry and may exploit information from the inherent structure. Here, it should be considered that terminological items – just as lexical items – show different degrees of complexity in their signifiers.

Characteristic 3. Different degrees of complexity of signifiers means that terms can range from morphologically simple forms like \textit{safe} to morphologically complex forms (e.g., \textit{safety}, \textit{Sicherheitsintegritätslevel}) or even syntactically complex forms (i.e., multi-word expressions like \textit{safety requirement}, \textit{equipment under control}). For stylistic reasons, compounds and multi-word expressions may show variation in complexity when used in texts, so that – apart from formally unrelated synonyms – complex items may have closely related signifiers as textual synonyms. For example, the English multi-word expression \textit{security level} may be realized as \textit{level of security} within a text. The same is true for compounds in languages like German where compounds like \textit{Sicherheitsintegritätslevel} can be realized as \textit{Level der Sicherheitsintegrität}. An algorithm should take this kind of variation into account, to learn adequate semantic representations for lexical or terminological items from syntactical words in corpora. A possible way to address this issue is to treat multi-word expressions and compounds as a bag of words and to aggregate their representations from their constituents’ embeddings. Overall, this should maximize the number of contexts semantic representations are learnt from. This approach is often taken when creating semantic representations for lexical items \cite[e.g.,][]{Arora.2017, Wieting.2016}. However, this should result in less valid lexical semantic representations, since morphologically and syntactically complex lexical items are semantically different from their constituents \cite[cf.][]{Klos.}. Accordingly, they have different contexts than do their constituents. This is most apparent in opaque compounds (e.g., \textit{jailbird}) or idiomatic expressions (e.g., \textit{to kick the bucket}). The context of both items should rather be different from that of their constituents \textit{bird} and \textit{bucket}. The same applies to a vast range of other items in the vocabulary with much higher degrees of transparency than the two aforementioned examples – look at \textit{tree house} or \textit{electric chair} respectively \cite[cf.][]{Herbermann.2002}. As regards domain-specific lexical items, we find, for example, \textit{trojan horse}. Aggregating embeddings for multi-word expressions from their constituents’ embeddings therefore does not rely on contexts that are relevant for the whole expression or compound but on those of their constituents. For the sake of validity and accuracy, the number of contexts that are relevant for the whole expression needs to be maximized. This is possible by treating multi-word expressions and compounds as one token. This way, only those contexts are permitted for learning the representations that are relevant to the whole multi-word-expression or compound. This, however, requires detailed knowledge about their variants to find \emph{all} relevant contexts, induces a greater need for a large domain-specific training corpus and the method possibly falls short compared to the aggregation method if such a corpus is not available.

Characteristic 4. As discussed in section \ref{sec-1}, terms are prone to polysemy – just as are all other lexical items. The number of related readings for technical terms, however, should be smaller than for everyday lexical items. Gladkova/ Drozd 2016 describe polysemy as “the elephant in the room” for intrinsic evaluations of word embeddings. Since evaluation data sets for lexical items are rarely controlled for polysemy or homonymy by providing raters disambiguating contextual information, they may not be able to evaluate the representation of a word properly. A representation may have been trained on a corpus where one sense of the word has been predominant, while raters of the evaluation data set predominantly favored a different sense.\footnote{A sense is usually considered to be one of the meanings a (lexical) word can have. For example, the (lexical) word \textit{house} is conventionally associated (among others) with the senses ‘building for human habitation, especially one that consists of a ground floor and one or more upper storeys’ and ‘noble, royal, or wealthy family or lineage’ (cf. \href{https://en.oxforddictionaries.com/definition/house}{https://en.oxforddictionaries.com/definition/house}, 2018-05-28).} For example, human raters may judge (\textit{apple}, \textit{pear}) to be more similar, while a semantic model may judge (\textit{apple}, \textit{computer}) to be more similar – depending on the training corpus.\footnote{Representations trained from very large multi-domain corpora should respectively rate both pairs as more or less equally similar.} The question for Gladkova/ Drozd 2016 is then, whether polysemous words should be removed from evaluation data sets, whether several senses should be provided to the raters, or whether semantic representations should account for polysemy and be sense-specific. They argue that they do not have to be sense-specific, since separate senses are just an expression of lexicographic convention. We, on the other hand, argue that to capture sense-specific representations would improve the validity of vector spaces that are supposed to be semantic spaces and not just spaces of formal words.

Of all the aforementioned characteristics of terminological entries, we will address each to a certain degree. Characteristics 2 and 3 will be addressed by using a weighted sentence embedding model, since a terminological entry is an entity that (like a sentence) consists of several lexical elements. It will treat a terminological entry like a bag of words and create aggregate embeddings from the single constituents of this bag of words. With this approach, we compensate for terminological variation and maximize the number of contexts, but not their relevance. Furthermore, we will adjust the embedding algorithm, which we hope will help us to account for characteristic 1 (see section \ref{sec-4.4}). As is the case with multi-word expressions, the disambiguation of vector representations is not yet state of the art, which is why we will not account for the fact that our sentence embedding model is based on polysemous word embeddings (see characteristic 4). The polysemy of the word vectors is partially compensated for by creating vector representations for terminological entries consisting of their term(s) and definition. Since a definition serves to clarify the sense of a term, a sentence embedding model thus creates sense-specific vectors by compensating for lack of relevant context. Therefore, even though our aggregate embeddings are “contaminated” with the polysemy of the word vectors to a certain degree, implicit disambiguation should be possible to some extent when the embedding input consists of more than just the terms of a terminological entry. The definition is furthermore provided to our raters so that they should favor the same sense as the algorithm or something more or less close to it.

\subsection{Sentence embedding model}
\label{sec-4.2}

We used the model described by \citet{Arora.2017}: a simple yet high-performing sentence embedding model that takes advantage of principal component removal and a straightforward and flexible weighting strategy. The choice of an appropriate model was determined by multiple parameters, which fall into three broad categories: 
\begin{enumerate}
\item the applicability to our data,
\item the practicality within the intended field of use,
\item the proven performance in semantic similarity estimations.
\end{enumerate}

First, though terminological entries refer to singular concepts, our data is nonetheless complex terminological entries (cf. section \ref{sec-4.1}), consisting of one or multiple synonym terms, compounds or term phrases as well as a definition comprising tokens of technical terms and general words alike. Within the entry, each element (synonym or definition) is a different expression of the same concept. Thus, it seems adequate to aggregate semantic information from these elements to a single embedding per entry that captures the referred concept in the embedding space.

For this reason, the simplest solution is to discard the structure of the terminological entry and assume all its tokens to form a "sentence" which can be captured by a sentence embedding model, treating all tokens as a single bag of words. For each token, a pre-trained word embedding is fetched and weighted according to the smooth inverse frequency (SIF) method proposed by \citet{Arora.2017}. It defines the weight for each word (or more precisely each word form or type) as

\begin{equation}
weight= \frac{a}{a+p(w)}
\end{equation}

\noindent wherein $a$ is a weighting parameter regulating the effect of the weighting scheme and $p(w)$ is the probability of a word as estimated from frequencies in a natural language corpus. Commoner lexical words, e.g., stopwords, are assumed to contribute less distinctive information, while rare words should be more specific and thus more important. The weighted average vectors are then modified by removing their projections on a number of top principal components $(d_{PCR})$, eliminating the greatest portion of variance within the embeddings that tends to be caused by more common and less important words as well. Both measures address sources of irrelevant semantic noise within the data set by different means and thus overlap partially.

Second, to support terminology harmonization in standardization, we prefer unsupervised approaches to semi- or fully supervised methods, since they neither require domain knowledge as input nor complicated and time consuming training techniques as proposed, for example, by \citet{Wieting.2016} (cf. also approaches discussed in section \ref{sec-2}). For our use case, it will be necessary to apply the same method to different domains. Supervised tuning steps for each domain or each project within different domains would therefore be a serious hindrance to adaptation. We therefore use the traditional 300-dimensional GloVe word vectors trained on the 840 billion token Common Crawl Corpus as a basic resource for our sentence embeddings (cf. \citealt{Pennington.2014}).\footnote{The GloVe word embeddings are publicly available at \href{http://nlp.stanford.edu/projects/glove}{http://nlp.stanford.edu/projects/glove}.} Several experiments by \citet{Arora.2017} with various corpora for deriving word probabilities yielded robust results, suggesting that using different corpora to estimate word probabilities does not necessarily lead to more adequate embeddings that fare better in the same task.\footnote{Those corpora were \textit{enwiki}, \textit{poliblogs} \cite{Yano.2009}, \textit{Common Crawl} \cite{Buck.2014} and \textit{text8} (URL: \href{http://mattmahoney.net/dc/textdata.html}{http://mattmahoney.net/dc/textdata.html}, accessed on July 12, 2018).} A domain-specific corpus, however, should emphasize some ubiquitous technical terms as most frequent ones in addition to functional words in general corpora. We therefore experiment with three different corpora, two of which are domain-specific. Due to a lack of availability of large corpora in our domain, we used our standards corpus and our entry corpus in comparison to \textit{enwiki}. An advantage of using such corpora is that they can be easily provided for any domain in standardization.

Third, despite having been coined as "embarrassingly simple", the method of \citet{Arora.2017} outperforms sophisticated model architectures on a variety of 22 textual similarity tasks. They apply it to one unsupervised (GloVe, cf. \citealt{Pennington.2014}) as well as one semi-supervised pre-trained embedding set (PSL, cf. \citealt{Arora.2017,Wieting.2016}). Both treatments were then compared to untreated embeddings (avg-GloVe, tfidf-GloVe, avg-PSL) as well as to embeddings from several supervised (PP, PP-proj, DAN, RNN, iRNN, LSTM, cf. \citealt{Arora.2017}) and unsupervised (skip-thought) models. While it is most competitive when applied to semi-supervised embeddings (being able to beat all other compared approaches in 12 out of 22 tasks), its application to GloVe beats the other unsupervised models in 18 out of 22 tasks (being itself beaten by the application to the semi-supervised approach in 11 of these cases). Using GloVe, \citet{Arora.2017} could show that the weighting by smooth inverse frequency improves performance by about 5\% as compared to unweighted averages. A 10\% performance gain was achieved by removing the first principal component from the matrix of sentence vectors, eliminating the component with the greatest variance. Using both yielded 13\% performance optimization.

\subsection{Sentence embeddings and evaluation method}
\label{sec-4.3}

In order to apply Harbsafe-162, we correlate the human ratings with the similarities of the SIF sentence embedding model. We computed entry embeddings, as analogous to sentence embeddings, according to the smooth inverse frequency (SIF) method by \citet{Arora.2017}, using our manually developed data set (cf. section 4) for evaluation. We applied the method with GloVe vectors to all terminological entries from our entry corpus (446 entries) and observed the association between cosine similarities and expert ratings from comparison pairs within Harbsafe-162.

\noindent In order to optimize our model, we studied the effect of variation of three factors:

\begin{itemize}
\renewcommand{\labelitemi}{$\--$}
\item The weighting parameter $a$,
\item different word probability distributions estimated from frequencies in various corpora, and
\item a varying number of top principal components to be removed.
\end{itemize}

Treatments 1 and 2 relate to smooth inverse frequency weighting and are referred to as W, while Treatment 3, concerned with principal component removal, is referred to as R. We will use this notation in our evaluation and discussion as well as in figures representing our results. To measure the improvement by the SIF-method we use unweighted averages as a baseline. First, by scaling the weighting parameter $a$ $(10^{-6}<a<10^{-1})$ it was possible to discern optimal performance as well as applicability and robustness of the method. Second, regarding the probability distributions, it seemed promising to evaluate the performance of estimated word probabilities $p(w)$ from domain-specific corpora that are easy to provide for our application in standardization. Thus, we test a distribution derived from our standards corpus. After removing the generic parts (header, footer, and administrative information of the standards), it consists of 196,216 tokens and ranks terms like \textit{security}, \textit{system}, and \textit{control} among the twenty most frequent terms. We refer to the resulting probability distribution as "standards probabilities" ($p_{standards}$). Furthermore, we evaluated the probabilities of the smallest set reasonably imaginable: a weighting scheme derived from our entry corpus itself (referred to as "entry probabilities"; $p_{entries}$). For comparison we used the far larger and more general \textit{enwiki }data set (“\textit{enwiki} probabilities”; $p_{\emph{enwiki}}$) that performed on par with \textit{poliblogs} \cite{Yano.2009}, \textit{Common Crawl} \cite{Buck.2014} and \textit{text8} in \citet{Arora.2017}. Third, we test the removal of different numbers of top principal components for the entry embedding matrix, $d_{PCR}\in\{0,…,6\}$.

To investigate further into the specifics of our data set, we also varied the composition of our sentence data. Besides full terminological entries, we also evaluate the performance of embeddings created from either the term tokens or the definition tokens of each entry. The term input contains one or several terms and/or term phrases per entry and therefore substantially fewer functional words (e.g., \textit{to}, \textit{in}, \textit{while}) than a definition. On average, term embeddings are based on two and definition embeddings on 22 tokens. We will denote the resulting vectors for full terminological entries as “entry embeddings” and the others as “term embeddings” or “definition embeddings” respectively.

We present our evaluation results by measuring the correlation of the cosine similarity between entry embeddings and median human ratings on our specified scale via Spearman's rank correlation coefficient $\rho$, satisfying requirements of our non-continuous rating scale by using ranked values for correlation.

\subsection{Results of evaluation with Harbsafe-162}
\label{sec-4.4}

Overall, we found an encouragingly strong correlation (up to Spearman’s $\rho=0.83, p<0.001$) between estimations of cosine similarity for pairs of entry embeddings and the corresponding expert ratings from Harbsafe-162 for a variety of configurations (cf. Table \ref{table6}). Regarding the effects of word weighting and top principal component removal, we were able to reproduce the findings of \citet{Arora.2017} with Harbsafe-162.

\begin{table}[h]
\caption{Maximal correlation coefficients (Spearman’s $\rho \times 100$) using individual optimal weighting parameters $10^{-4} < a < 10^{-2}$ for various inputs and variations of the SIF-model. The model appears to be robust to different weighting schemes (W) and benefits greatly from principal component removal (R) within embedding sets. For each result marked R, one component was discarded ($d_{PCR}=1$).}
\resizebox{0.75\textwidth}{!} {\begin{minipage}{\textwidth}
\renewcommand{\arraystretch}{1.5}
\begin{tabular}{lcccccccccc}
\hline
\multirow{2}{*}{} & \multicolumn{2}{c}{\multirow{2}{*}{\textbf{Unweighted averages}}} & \multicolumn{2}{c}{\multirow{2}{*}{\textbf{$p_{\emph{enwiki}}$}}} & \multicolumn{2}{c}{\multirow{2}{*}{\textbf{$p_{standards}$}}} & \multicolumn{2}{c}{\multirow{2}{*}{\textbf{$p_{entries}$}}}\\
\multicolumn{1}{p{2cm}}{}& \multicolumn{1}{p{1,5cm}}{}& \multicolumn{1}{p{1,5cm}}{} & \multicolumn{1}{p{1,5cm}}{} & \multicolumn{1}{p{1,5cm}}{} & \multicolumn{1}{p{1,5cm}}{} & \multicolumn{1}{p{1,5cm}}{} & \multicolumn{1}{p{1,5cm}}{} & \multicolumn{1}{p{1,5cm}}{} \\
 \cline{2-9}
\textbf{Token Input} & \multicolumn{1}{c}{\textbf{--}} & \textbf{R} & \textbf{WR} & \textbf{W} & \textbf{WR} & \textbf{W} & \textbf{WR} & \textbf{W}
\\ \hline
Entries & 67 & 80 & \textbf{82} & 74 & \textbf{83} & 74 & \textbf{83} & 72
\\  \hline
Definitions & 61 & 74 & 78 & 69 & 78 & 71 & 78 & 68
\\ \hline
Terms & 67 & 77 & 77 & 68 & 78 & 68 & 78 & 68
\\ \hline
\end{tabular}
\end{minipage} }
\label{table6}
\end{table}

The weighting scheme had the biggest effect on definitions and a neglible effect on terms, supposedly due to the relative lack of common words in technical terms and term phrases. For entry and definition embeddings, it contributed 5-7\% by itself (W) as compared to the baseline and 2-4\% in combination with principal component removal (WR-R). Both measures aim at reducing the detrimental, non-distinctive effect caused by probable and common words, which tend to contribute most strongly to overall variance. Insofar, both measures overlap but still have the greatest effect when applied together.

\begin{figure}[h]
\centering
\includegraphics[scale=1]{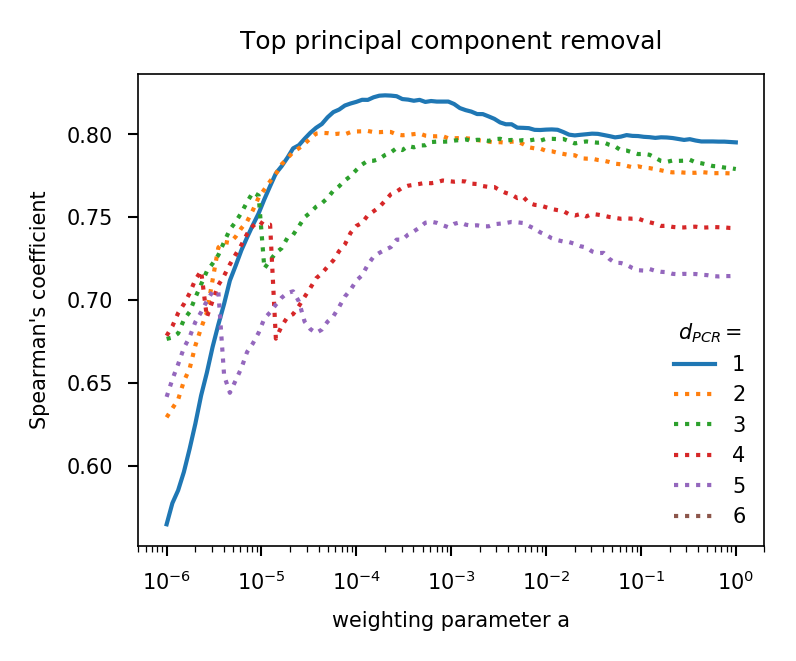}
\caption{Effect of weighting parameter $a$ using the \emph{enwiki} probabilities for entry embeddings for a range of top principal components to be removed.}
\label{fig.6}
\end{figure}

With entry and definition embeddings, top principal component removal by itself (R) had an impact of 13\% compared to the baseline, while term embeddings facilitated a smaller but still sizeable improvement of 10\%. In combination with weighting (WR), the improvements from principal component removal are diminished to 8-11\% (WR-W). Generally, we found the removal of only one top principal component to be optimal in terms of both performance and stability (cf. Figure \ref{fig.6}).

\begin{figure}[h]
\centering
\includegraphics[scale=0.85]{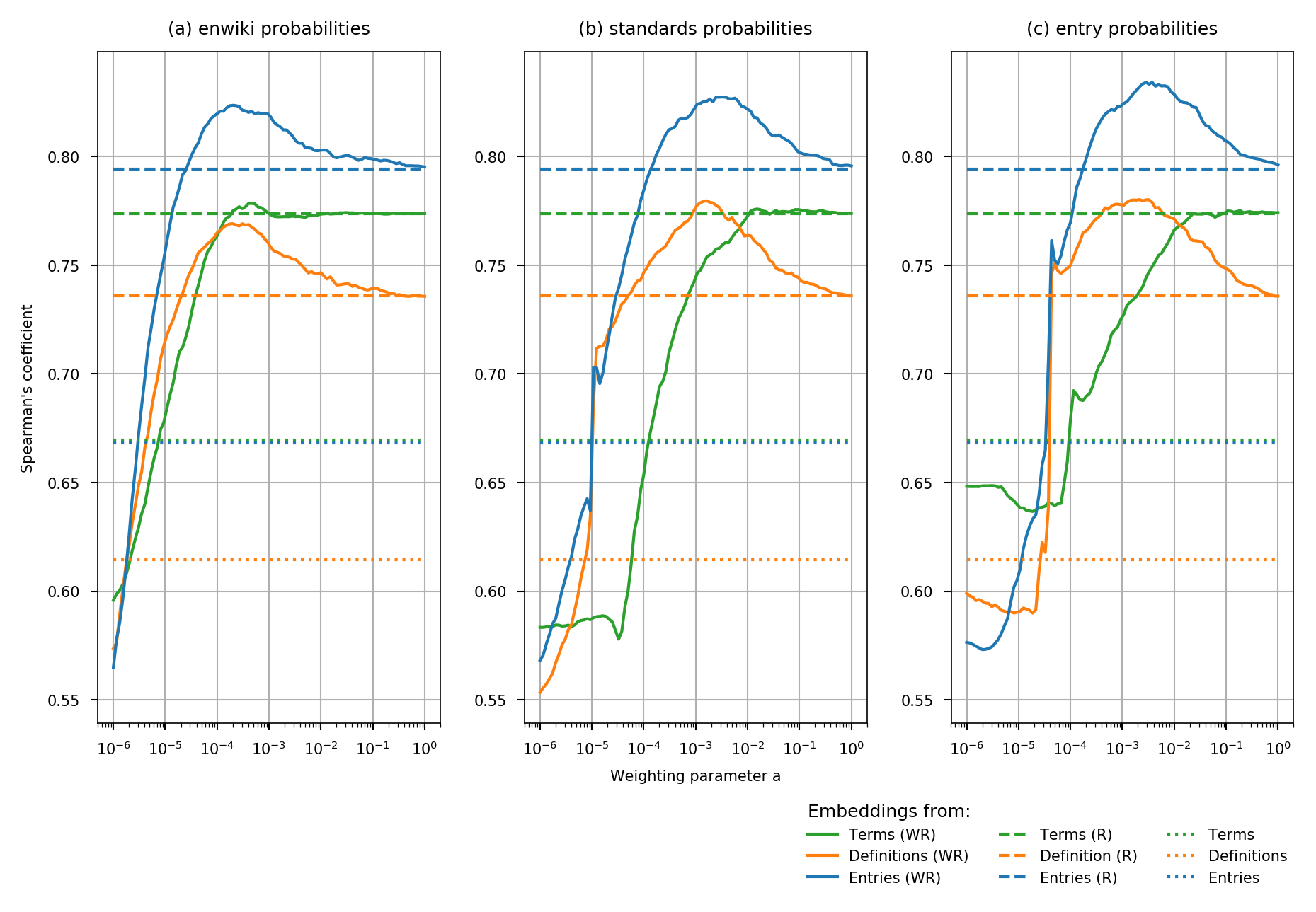}
\caption{Effect of weighting parameter $a$ within the SIF-model (WR) across different inputs compared to corresponding results of unweighted averages with and without principal component removal (R, $d_{removed}=1$).}
\label{fig.7}
\end{figure}

The modest improvements facilitated by the weighting scheme proved to be surprisingly consistent across different estimations of word probability derived from extremely dissimilar corpora. Optimal performances using each corpus varied within 1\% (cf. also Figure \ref{fig.7}). Other than hoped for, there was no marked improvement in overall performance caused by domain-specific word probability distributions, but neither did performance deteriorate, which might have been expected just as well, due to possible exaggerations in the rather small word frequency corpora. Instead these resulted merely in a requirement to apply a greater weighting parameter $a$, scaling down the weighting scheme’s influence, while causing some instabilities at $a< 10^{-4}$ (cf. Figure \ref{fig.7}b and \ref{fig.7}c). This effect is exacerbated for term embeddings that heavily depend upon probable domain-specific words (as in \textit{security patch} or \textit{safety integrity}). For the enwiki probabilities, however, performance peaked for weighting parameters where $10^{-4}<a<10^{-3}$, reproducing the findings of \citet{Arora.2017} (cf. Figure \ref{fig.7}a).

While cosine similarity between entry embeddings can be established as the best predictor for human similarity estimations of terminological entries, cosine similarity of weighted definition embeddings competes with that of term embeddings, marginally taking the lead with a domain-specific word frequency corpus (cf. Figure \ref{fig.7}b and c). This suggests that, while there is no explicit/ direct lexical signifier in the definition that references the concept described by a terminological entry, the surplus of encoded information in the aggregate vector representation compensates for lack of specificity on a token by token basis.

\begin{figure}[h]
\centering
\includegraphics[scale=.875]{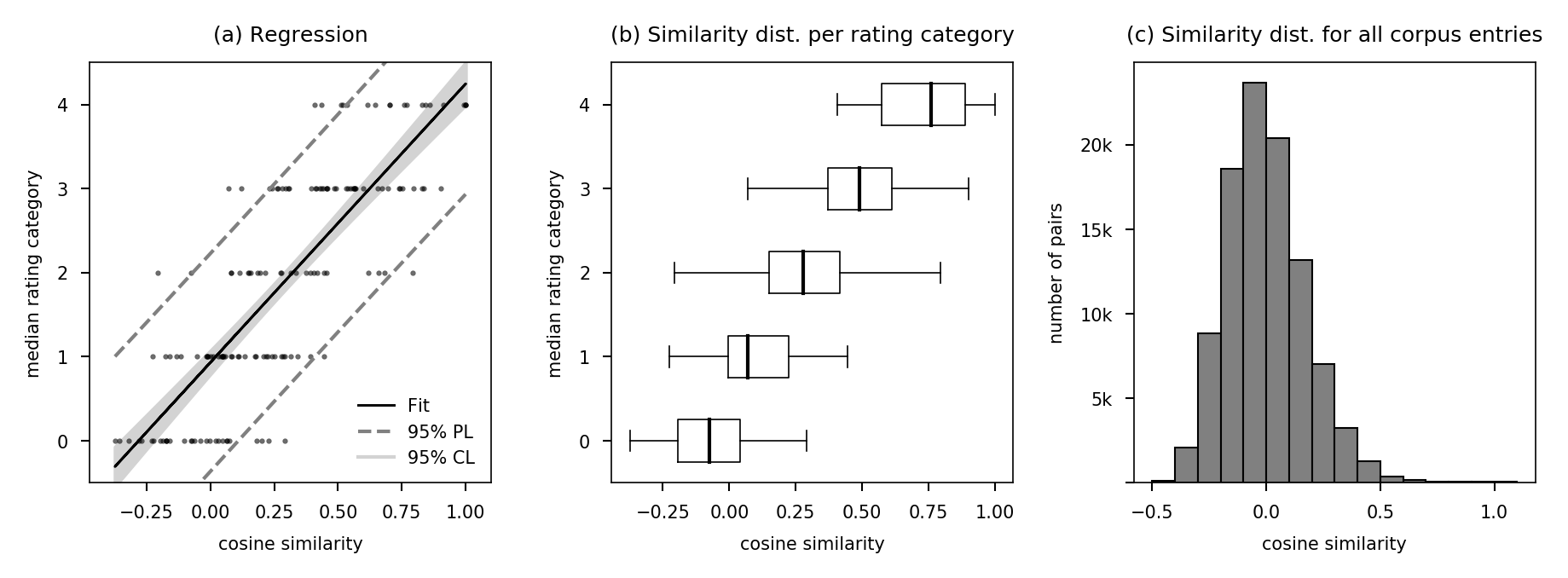}
\caption{a) Cosine similarity between entry embeddings in relation to median rating category (five raters) with 95\% predictive limitis (PL) and 95\% confidence limits (CL), produced with the terminological entry corpus weighting scheme ($a = 0.0038$), first principal component removed (Spearman $\rho = 0.83$). b) Box plot (same model): distribution and outlier in category 4 may be falsely assigned due to heavy skew in input data. c) Distribution of the similarities of all entry combinations within the entry corpus.}
\label{fig.8}
\end{figure}

Depending on the use case, the predictive power of our model should not be overstated, even though it is quite strong in correlation. For example, to predict all individual pairs within category 0 that remain within the 95\% prediction interval a reasonable cut-off point for cosine similarity $s$ would have to be set at $s< 0.1$ (cf. Figure \ref{fig.8}a and b). This, however, would also select false positives, that is, pairs from categories 1, 2, and 3, while at the same time not selecting all observed pairs of category 0. To capture all pairs from category 0, a cut-off point at $s< 0.3$ would be required. This increases true positives and, unfortunately, also the number of false positives from other categories, while both cut-off points would select more than half of the population (cf. Figure \ref{fig.8}c).

The assessment of the validity of the rating scale and Harbsafe-162 showed a lack of non-trivial category 4 pairs. Instead, Harbsafe-162 contains mostly trivial category 4 pairs (i.e., pairs sharing most if not all of their tokens) which causes a heavily skewed distribution of cosine similarities for category 4. Those trivial pairs act as an outer bound near the center, while they are actually better suited to form the long tail of the distribution. Some identical concepts that are either denoted by different terms or defined differently (or both) were found at $0.9< s< 0.7$, where the true median similarity of pairs within category 4 may be suspected. This has likely lead to a slight overestimation of correlation. Further conclusions and mitigating measures will be discussed below.

\section{Discussion of the experiment}
\label{sec-5}

The evaluation of our model sentence embedding method by \citet{Arora.2017} (SIF) has succeeded in (1) reproducing established findings about the method at hand within our scientific and technical domain with its specific data structure, (2) exposing some shortcomings within the evaluation set, and (3) showing applicability of the method for our intended use case.

First, the SIF method which had been shown to be a robust, flexible, easy to use and yet competitive state-of-the-art baseline by \citet{Arora.2017}, has also been confirmed for our technical and scientific domain. Unsupervised general language word embeddings, while not optimized for the use within a specific scientific and technical domain, still proved serviceable in predicting the similarity of highly technical concepts. Even though some reservations due to comparative modesty in size and the small number of raters (however expert they may be) seem appropriate, the strong correlation between predicted similarity and expert judgements can be regarded as proof of concept.

Furthermore, our straightforward assumptions regarding the possibility of simplifying the structure of our data to simple bags of words turned out to be at least serviceable if not adequate. Combining one to multiple terms and term phrases with the definition improved correlation with our expert ratings by a noticeable amount over term embedding averages ($\rho_{Terms}=0.78$; $\rho_{Entries}=0.83$). Surprisingly, weighted term embeddings did not perform better than definition embeddings even though they are based on a direct lexical signifier for the concept, while the definition is only a paraphrase of the concept. On the one hand, it would be promising to optimize different compositions of entry embeddings by applying an additional weighting step. There is no inherent reason why tokens from terms and tokens from definitions should effect the entry embedding to the same degree. A weighting parameter may be considered to optimize the proportions against the evaluation set. On the other hand, it seems likely that SIF-weighting for term tokens is wholly omissible within entry embeddings, since no positive weighting effect could be shown for term embeddings. A weighting parameter for optimizing the effect of tokens from definitions might suffice.

As regards the question whether the technical language of a domain may be modeled more appropriately by using inverse weighting by domain-specific word probabilities, the experiment was not conclusive. While it might be argued that our entry corpus and standards corpus were too small to derive a good estimate of a probability distribution from, the steadiness of the results may also suggest that any probability distribution would be serviceable as long as it ranks semantically insignificant functional words near the top. Focusing on individual data points, however, the aggressive down-weighting of domain-specific ubiquitous words like \textit{security} or \textit{system}, has a profound impact on the residuals between measured and expected similarities as predicted from median ratings. For some pairs, \emph{enwiki} probabilities seem more appropriate, while for others, those derived from our standards and our entry corpus. We cannot report what exactly causes that. Generally, the \emph{enwiki} probabilities weigh down common functional words more strongly by one order of magnitude. A straightforward omission of traditional stopwords could compensate for that and possibly improve the use of small domain-specific word frequency corpora beyond more general predictions of word probability, which barely distinguish between most technical terms.

Second, regarding shortcomings in the evaluation set, the biggest problem from an experimental point of view were the trivial pairs in category 4. While composing the evaluation set we found too few non-trivial examples of entries that refer to the same concept without being literally identical or very close to it – a phenomenon that is highly undesirable but quite common in standardization. We consequently lack the means to set similarity thresholds for the prediction of those entries. Furthermore, our inclusion of near-identical entries most likely leads to a slight overestimation of correlation and possibly to an overemphasis of token-by-token identity instead of conceptual identity when using Harbsafe-162 for measuring performance. Thus, replacing and extending pairs within category 4 is a priority and will require the inclusion of additional entries from standardization sources that are external to our standards corpus.

Third, our results may be regarded as an important first step with respect to the applicability of embedding methods within our intended use case. The prediction of similarity of terminological entries, however, does not provide quick and easy solutions for the challenges of terminology work by itself. The initial success needs additional efforts for practical use.

Regarding the resolution of doublettes within standards or within a set of standards (cf. section \ref{sec-1}), requirement \ref{req6} cannot yet be modeled in a quantifiable way, because of the missing evaluation data for category 4 containing highly similar pairs. After this deficit is being resolved, we expect a decent reduction of candidate pairs while preserving high recall and very few false positives at the top of the ranks. Requirement \ref{req7} is profoundly harder to solve. According to our initial consistency model, the reduction of relevant and thus mandatory consistency checks between terminological entries is vital to help experts assure terminological consistency within a standard. Filtering mandatory checks could have been easily automated, had it been possible to differentiate concepts without any relation to each other (category 0) from the (presumably smaller) number of entry pairs that are important to check for consistency (categories 1-3, cf. section \ref{sec-1}). As it turns out, however, pairs from category 0 largely overlap with category 1 and even 2 in predicted similarity (cf. Figure \ref{fig.8}b). On this basis, only a very small portion of the population can be safely excluded for irrelevancy. The consequent reduction of recall that would likely be inherited by any down-stream application must be addressed through various additional measures depending on more narrowly defined use cases.

Wherever high recall is needed (as is the case for consistency checks), common sources of inconsistencies must be further defined and predicted via additional features. A simple example would be pairs where the term similarity is high, indicating high relevance for consistency, while the definition similarity is comparatively low. Those concept pairs are likely defined in an incoherent and possibly contradictory manner. Definitional dissimilarity may be very well suited to serve as a priority ranking for the resolution of doublettes as well. We will explore this and other avenues further.

Where high recall is negligible, however, automatic semantic assistance in the standardization process may still benefit directly from our results. Experts can be provided with automatically selected, highly relevant pairs of items from different standards (whose overlap is not necessarily known to their respective standards bodies) and may thus gain a higher coverage of consistency checks. In addition, any editorial decision can be assisted by traditional information retrieval methods. It is a common practice to inform decisions regarding one item by providing context through similar and thus potentially relevant items. Of course, such relevancy rankings for each entry need accuracy evaluation, but preliminary samples are already appealing at least (cf. Table  \ref{table7}). Visualization and clustering are promising as well.

\begin{table}[h]
\caption{Cosine similarity rankings for entries based on entry embeddings. Duplicate terms are removed.}
\resizebox{0.75\textwidth}{!} {\begin{minipage}{\textwidth}
\renewcommand{\arraystretch}{1.5}
\begin{tabular}{cccc}\hline
\textbf{risk (1)}&\textbf{firewall}&\textbf{bug}&\textbf{encryption}\\ \hline
risk mitigation & router & vulnerability & decryption \\ \hline
likelihood & safety network & patch & plaintext \\ \hline
consequence & field I/O network & risk (2) & secret key \\ \hline
impact & PPP & threat & symmetric key \\ \hline
harmful event & gateway & countermeasure & ciphertext \\ \hline
risk assessment & network layer protocol & software safety integrity & symmetric key algorithm \\ \hline
hazard & untrusted channel & hardware safety integrity & public key \\ \hline
security incident & host & fault tolerance & cryptographic algorithm \\ \hline
dependent failure & demilitarized zone & proof test & pseudorandom number generator \\ \hline
harm & conduit & reasonably foreseeable misuse & compromise \\ \hline
risk assessment & local area network & dynamic testing & asymmetric key algorithm \\ \hline
\end{tabular}
\end{minipage} }
\label{table7}
\end{table}

\section{Discussion of the rating task}
\label{sec-6}

We would like to give a critical review of our own rating method, since there is an ongoing controversy about intrinsic evaluation methods and the refinement of data sets, which is mainly concerned with their quality and validity (cf. for example the contributions in the Proceedings of first Workshops on Evaluating Vector-Space Representations for NLP \cite{Bowman.2016, Bowman.2017}). Consequently, alternative means of evaluation are suggested which are considered extrinsic: \citet{ Batchkarov.2016}, for example, argue that the “quality of an unsupervised algorithm can […] only be assessed in the context of an application” \cite{Batchkarov.2016}. Since harmonization of terminological entries as described by requirements \ref{req6} and \ref{req7} is mainly concerned with finding highly similar terminological entries, our assessment of the embeddings with an intrinsic evaluation method is rather justified even though our approach may have inherited issues simply by adopting the rating task methods of \citet{CamachoCollados.2017}. The following section, therefore, serves the purpose of giving a short introduction into this debate and assessing in how far these arguments apply to Harbsafe-162.

\citet{Hill.2015} criticized a range of lexical data sets (e.g., WordSim-353) for not properly distinguishing between similarity and relatedness. They observed word comparison pairs that receive high ratings on a similarity scale merely because they are related, not because of their similarity, e.g., (\textit{musician}, \textit{microphone}). In some cases, these ratings were even higher than ratings for truly similar pairs, thereby confusing categories that ought to be distinct. This, of course, has an effect on the evaluation of distributional semantic models and their predictive power. Embedding models that put related and similar pairs near each other when they are supposed to separate them, do not receive enough punishment from evaluation data sets that do not properly distinguish between similarity and relatedness either. Our own requirements for harmonization make it necessary to properly distinguish between relatedness and similarity. Requirements \ref{req1}, \ref{req2}, and \ref{req3} are best addressed by similarity, while the consistency requirement relies on both. Thus, the distinction between similarity and relatedness is relevant for Harbsafe-162. We will therefore give a short introduction of the issues described by \citet{Hill.2015} and, with respect to this debate, evaluate Harbsafe-162 for its reliability regarding the distinction of similarity and relatedness. After that, we will join a more general discussion of the design of rating tasks and intrinsic evaluation data sets from a point of view of lexicology and terminology science.

\citet{Hill.2015} blame the confusion between relatedness and similarity observable in rated pairs on the instructions provided to the raters by most rating tasks. These often do not properly distinguish between similarity and relatedness and rather use both terms almost like synonyms. In an attempt to create a new lexical evaluation data set (SimLex-999), they accordingly altered the instructions to make sure that raters are able to tell the difference between relatedness and similarity.\footnote{This, however, does not fix another issue: Similarity is a very diverse concept and similarity between two items can be stated on very different grounds. Two lexemes can be considered similar for a multitude of reasons. They may be in a specific kind of relation (\textit{cat}, \textit{pet}); they share the same referent (\textit{Gwen Stefani}, \textit{singer}); they share connotational values (\textit{gas guzzler}, \textit{clunker}); their referents are physically similar (\textit{glass tabletop}, \textit{window pane}); their linguistic forms are similar (\textit{meronymy}, \textit{metonymy}) or they are functionally similar in a larger frame of items (\textit{term}, \textit{word}). This is a further source of unreliability of evaluation data sets that should be further explored.} They phrased the instructions in a way that more suitably conveys the contrast between relatedness and similarity by stating that related words “are not […]\footnote{We left a second \textit{not} out of this quote, which can be found in the article. This double negation must be an error in \citet{Hill.2015} which hopefully did not occur in their rating tool.} very similar” or “not necessarily similar” \cite{Hill.2015}. The instructions should, therefore, implicitly give raters the idea that one pole of the rating scale denotes the absence of similarity, even though participants would still have to rate both on one scale. Even though these efforts result in higher inter-annotator agreement and thus higher reliability as compared to other lexical data sets, there are still some issues with their rating procedure as regards similarity and relatedness. We will shortly come back to these issues.

As regards Harbsafe-162, we chose a similar conception of the rating scale as \citet{Hill.2015} have. Unfortunately, we have not made the contrast between similarity and relatedness just as clear as \citet{Hill.2015} have – which a closer look at our rating scale reveals. The scale points are labeled by category names that are supposed to show a degree of similarity. The labels on our scale, however, are problematic, since not all of them fit the linguistic example data. Point 2 is especially problematic since it is labeled “slightly similar”. This label is simply not applicable to the example: Or how is a \textit{pilot} slightly similar to a \textit{plane}? Our rating scale therefore shows the criticized confusion of similarity and relatedness. Point 2 on our scale should rather have been named “dissimilar and directly related”, while point 1 should have been named “dissimilar and indirectly related”. Consequently, we should be able to find pairs that show this confusion of similarity and relatedness in Harbsafe-162, that is, dissimilar pairs that were rated as similar and some similar pairs that were rated as dissimilar. Identifying them must rely on an educated guess, for which we set the authors’ unanimous ratings as an estimation for the true category of a pair\footnote{We consider this a sufficiently valid decision, since we designed the scale with a certain purpose in mind and use it based on our experience in lexical semantics.}. All in all, we rated 134 pairs unanimously. For these we compare the estimation of the true value to the median rating of all other raters and look at the deviations (see cells in boldface in Table \ref{table8}).

\begin{table}[h]
\centering
\caption{Number of similar pairs (3-4) rated “dissimilar” and dissimilar pairs (0-2) rated “similar”}
\resizebox{0.75\textwidth}{!} {\begin{minipage}{\textwidth}
\renewcommand{\arraystretch}{1.5}
\centering
\begin{tabular}{cccccc} \cline{2-6}
\multicolumn{1}{p{3cm}}{} & \multicolumn{5}{c}{\multirow{2}{*}{\textbf{domain experts (median ratings)}}}\\
\multicolumn{1}{c}{} &  \multicolumn{1}{p{1cm}}{} & \multicolumn{1}{p{1cm}}{} & \multicolumn{1}{p{1cm}}{} & \multicolumn{1}{p{1cm}}{} & \multicolumn{1}{p{1cm}}{} \\ \cline{1-6}
\multicolumn{1}{c}{\textbf{authors (avg. rating)}} & \multicolumn{1}{c}{4} & \multicolumn{1}{c}{3} & \multicolumn{1}{c}{2} & \multicolumn{1}{c}{1} & \multicolumn{1}{c}{0} \\ \hline
\multicolumn{1}{c}{\textbf{3-4}} & \multicolumn{1}{c}{20} & 16 & \textbf{6} & \textbf{1} & \textbf{0}\\ \hline
\multicolumn{1}{c}{\textbf{0-2}} & \multicolumn{1}{c}{\textbf{2}} & \textbf{14} & 6 & 35 & 34\\ \hline
\end{tabular}
\end{minipage} }
\label{table8}
\end{table}

Based on average author ratings with a clear tendency towards the upper or lower end of the scale, we can single out seven pairs that are likely not rated similar enough (5\%) and sixteen pairs that are likely rated too similar (11\%, for a list of some examples see the Appendix in section \ref{appendix}). Overall, we estimate 73\% of all rated pairs in Harbsafe-162 to be reliable concerning the distinction between relatedness and similarity.

Let us now come to the more general discussion about designing and optimizing evaluation data sets. This requires a closer look at the remaining issues of the rating procedure by \citet{Hill.2015} that are mostly located in their rating scale. These have been criticized by \citet{Avraham.2016} who consequently suggest further methodical adjustments to optimize the reliability of data sets for intrinsic evaluation of distributional semantic models. The most important point of their criticism is the fact that \citet{Hill.2015} – despite their intention to differentiate between similarity and relatedness – still use only one rating scale for different relation types. The poles of this scale are labeled as “less similar” and “more similar” ranging from 0 to 6, with 0 being located on the “less similar” pole and 6 on the “more similar” pole. As you can see, this rating scale is contrary to that part of their rating instructions that emphasizes the contrast between relatedness and similarity. With every batch of word pairs, raters were shown this scale with the short instructive reminder to “[r]ate the following word pairs according to how similar they are”. Therefore, they may still have been misled – despite the optimized initial instructions for the rating task. SimLex-999 may accordingly still show unreliability as regards similarity and relatedness.

Furthermore, the instructions of \citet{Hill.2015} still lacked information on how to differentiate different degrees of relatedness so that pairs with related items receive an accurate rating, that is, one that is close to the desirable rating. According to \citet{Avraham.2016}, this lack leads to some considerably inconsistent ratings for comparable pairs rendering the values of the whole scale ambiguous in that they sometimes do indicate a difference between the relations in two rating pairs and sometimes not. Raters, in sum, treated comparable pairs differently, thus even SimLex-999 has reliability problems that affect evaluations of distributional semantic models. What could be the solution to this problem? \citet{Avraham.2016} make several more or less radical suggestions. One of them is, to create specialized evaluation data sets that focus on one single lexical-semantic sense relation type. Such a data set would still contain pairs with other relation types, but would treat them differently during its compilation. The relation type in the focus is called a “preferred” relation type, while all other relation types are unpreferred. Only pairs with the preferred relation will be part of a rating task, while pairs with unpreferred relations will be assigned a random value that is lower than the value of the lowest rating value. Giving raters only items with the same relation type should help to improve the reliability of a set, since raters are no longer asked to take part in two different rating tasks at the same time. Instead, they can focus on consistently comparable pairs. How would a rating task in this approach look like? Raters would have to use a scale to indicate the strength of the relation for each pair (strength, since similarity is not applicable to all relation types). \citet{Avraham.2016} demonstrate this approach by focusing on the relation types of hypernymy and co-hyponymy. These relation types are similarity-based and therefore a scale to measure their strength may legitimately be characterized as a graded scale for similarity. In other cases, this may not be appropriate. In fact, for some relation types even gradedness may not be applicable. We will soon come back to this point. When the data set is used to evaluate a distributional semantic model, the model should prefer the same relation type as the evaluation data set – that is, it should rank all pairs that show the preferred relation type at the top of the list. At the bottom of the list, by contrast, should be those pairs that show the unpreferred relation types.

While this is a good step towards better reliability, it is legitimate to ask whether it is really necessary to connect only one single relation type to a similarity scale and a similarity rating task. \citet{Hill.2015} have shown that several different relation types, which are, however, all similarity-based, take mostly distinct positions on a similarity rating scale. The order of relations (which would equal a succession from strongest similarity to weakest similarity) is synonymy > hypernymy (1)\footnote{The number behind the hypernym relation indicates the number of edges between the items in the relation.} > hypernymy (2) > hypernymy (4) > hypernymy (3) > co-hypernymy > meronymy > hypernymy (5)\footnote{We left antonymy out of this succession, even though it stands at the last position in \citet{Hill.2015}. For antonymy, the whole point is that there is opposition or incompatibility and similarity at the same time, e.g., (\textit{dead}, \textit{alive}). Antonyms should usually make annotators hesitate unless instructed otherwise. In the case of \citet{Hill.2015}, raters were not provided specific instructions on antonyms but were reminded to think of synonymy if ever they are uncertain – this could explain why antonyms received low similarity values in SimLex-999 and settled at the lower end of the scale. Possibly annotators stressed the oppositeness of items in antonym pairs and their overall dissimilarity to synonym pairs. Other rating tasks (e.g., Word-Sim 353) demand to rate them as similar which would probably put them somewhere else.}. This is very consistent with what one would expect from a purely theoretical standpoint: For synonymy, hypernymy, and co-hyponymy, similarity is a central notion. For meronymy (i.e., part/whole-relations), similarity plays a role only in some instances of the relation. Consider for example a table consisting of a tabletop and legs: For a massive wooden table, tabletop, and legs are similar insofar they are made of the same material; this similarity is however not given when the tabletop is made of glass and the legs of wood. If we were to distinguish further types of meronymy, these may show an expectable pattern of similarity/dissimilarity as well. In addition, if we were to add equivalence as a relation between multilingual data, it should settle with synonymy. The only astonishing ranking is that of hypernymy and co-hyponymy. Here, even pairs that have a distance of four edges are considered more similar than co-hyponyms, while pairs with five edges are rated a little lower. Co-hyponyms can be very similar, even more so than direct hypernyms – compare for example (dining table, desk) and (table, furniture). This, of course, depends on several factors, for example the level of abstraction, which can be different for different co-hyponym pairs like (dining table, desk) and (table, chair). We therefore blame co-hyponymy’s low rating on its comparatively undifferentiated treatment by \citet{Hill.2015}. While hypernymy has been characterized by distance, no such thing has been done to co-hyponymy, so that \citet{Hill.2015} compare apples to oranges. It would be possible, however, to apply metrics like the relative depth of a concept in a graph-based lexical resource to differentiate the co-hyponymy relation in a way that is meaningful for similarity ratings. Controlling for level of abstractness could help to make a data set more reliable – just like controlling for edge-wise distance between items in hypernymy pairs. Overall, this suggests two things. First, relation types may be processed within one and the same rating task with one and the same rating scale – as long as they apply the same criterion, which in this case is similarity, and as long as they settle on different positions on the rating scale. Below, we will provide further (methodological) support for this observation. Second, the modeling of relation types for a rating task needs to reach a sufficient and linguistically grounded degree of granularity to guarantee high reliability.

Let us now come back to the observation we made above: That the notion of similarity or even the notion of gradedness that is applied by rating scales, is not applicable to all relation types. It is, in fact, a question whether all relation types can be graded in the same way hypernymy and hyponymy can. Consider for example the German synonym pairs (\textit{Orange}, \textit{Apfelsine}) (en. ‘orange’ (n.)) and (\textit{Samstag}, \textit{Sonnabend}) (en. ‘saturday’ (n.)) – which of them is more synonymous (or similar) than the other? Is it even possible to distinguish a strength of the relation of truly synonymous pairs? At least, it would not be along the line of a graded scale but along the lines of distinct relation types, such as full synonymy and partial synonymy (for example (\textit{car}, \textit{vehicle}) cf. \citealt{2342.2011}).\footnote{Partial synonymy is a problematic relation because it can be defined very differently depending on the theoretical background: For some disciplines, like terminology, partial synonymy is something like a conceptual overlap – this would make quite a proportion of the vocabulary to partial synonyms \cite[cf.][]{Drewer.2017}. For structuralist linguists, partial synonymy may be a sub-lexical relation, that is, a relation of (full) synonymy between senses of different (polysemous) lexemes (as compared to complete synonymy as a relation between two lexemes (all senses included).} Accordingly, synonymy and partial synonymy are themselves degrees of similarity but should not (each) be subdivided for different degrees of similarity themselves. This has a further advantage: As \citet{Avraham.2016} themselves state, rating hypernyms on a similarity scale is not a trivial task and – to our opinion – should become even less trivial for synonyms. Using one ordered scale for several similarity-based relation types makes the rating task easier for the rater, since the rating scale would combine types that are more easily distinguishable. It is, for example easier to assign different values to pairs like (\textit{midday}, \textit{noon}) and (\textit{lion}, \textit{zebra}) than to pairs like (\textit{midday}, \textit{noon}) and (\textit{motherboard}, \textit{mainboard}).

The question whether gradedness is a relevant aspect of relation types, becomes even more important for relation types that only show relatedness, not similarity. While it is quite comprehensible to distinguish similarity-based relation types and to rank (all of) them on the same ordinal scale, this is not as intuitive for relation types that only exploit the notion of relatedness, even though we admit that it is intuitive to grade relatedness of pairs like (\textit{cup}, \textit{coffee}), (\textit{software}, \textit{keyboard}), and (\textit{pencil}, \textit{frog}). According to the latter intuition, the rating scale of \citet{CamachoCollados.2017}, as well as our own rating scale, reflects a ranking of relatedness. Let us explore this further: As regards gradability of single relation types of this kind, we have to admit that they sometimes may build chains that are comparable to hypernym chains. Just like in hypernymy, the “edges in between” may be a factor determining different degrees of strength in such relations. Consider for example the following temporal relations: Which is stronger – (\textit{Monday}, \textit{Tuesday}) or (\textit{Monday}, \textit{Wednesday})? Now, on the other hand, compare the pairs (\textit{Apple}, \textit{iPhone}) and (\textit{Amazon}, \textit{Fire Phone}) that are in a relation of origination. Theoretically, there should not be a strength of this relation. The relation just is, or it is not. However, there may still be effects resulting from relevance or familiarity that strongly depend on the raters’ world knowledge. For example, a rater may be uncertain whether the relation of the pair (\textit{Kia}, \textit{Ceed}) is equally true as the relation between (\textit{Volkswagen}, \textit{Golf}). However, this is not a difference in degree of the relation type itself. As regards a relatedness scale accommodating several types of relatedness, the problem is that there is a vast number of such relation types (e.g., causality, meronymy, temporal relations, cf. \citealt{Nuopponen.2005}) and it is not clear or apparent which ones are stronger. Theoretical hypotheses might be provided by lexical field theory or frame semantics. The latter distinguishes between the core and the periphery of a frame (cf. \citealt{Busse.2016,Geeraerts.2010}). This question could also be answered empirically, though, by creating a data set focusing on relatedness instead of similarity. Until such a grounded ordered scale of relatedness is available, the approach to create separate evaluation data sets for single relatedness relation types, as given by \citet{Avraham.2016}, is a valid one – at least for methodical reasons concerning reliability.

For task-related reasons, on the other hand, this evaluation approach lacks validity in terminology harmonization. When evaluation is based on one relation type only, there needs to be a decision on which relation type needs to be featured by an evaluation data set. \citet{Avraham.2016} assume that this choice should be based on the down-stream application a distributional semantic model is supposed to support. A similar argument is brought forward by \citet{Hill.2015}. They consider evaluation data sets that focus on similarity to be especially important for lexical resource building. Here, we come back to our motivation as discussed in section \ref{sec-1}. Modern lexical resources go way beyond classical glossaries and try to be knowledge resources. This is especially so for domain-specific resources, where onomasiological modeling of terminology has become predominant. Lexical and terminological resources accordingly become multiply related data structures (for example \textit{iglos} \cite[cf.][]{Schnieder.2010}, Frame-based Terminology \cite[cf.][]{Faber.2014}, FrameNet \cite[cf.][]{Ruppenhofer.2016}, or WordNet \cite[cf.][]{Fellbaum.1998}). For the creation of such lexico-terminological resources, it is therefore not sufficient to create data sets solely based on one type of relation type, that is, similarity- or relatedness-based ones. For terminology work and terminology harmonization (cf. section 1), very different relation types – some from the relatedness spectrum, some from the similarity spectrum – play an equally important role. The consequence would be to create several evaluation data sets instead of one. We therefore have paid for validity and methodical efficiency with some limitations in reliability.

\section{Publication of Harbsafe-162 and future revisions}
\label{sec-7}
Harbsafe-162 has been published online at \href{https://www.dke.de/de/themen/projekte/harbsafe}{https://www.dke.de/de/themen/projekte/harbsafe}.
We plan to update Harbsafe-162 in two respects. First, we will provide an update called Harbsafe-172 to improve the quality of the ratings pairs, i.e. by including more non-trivial category 4 entry pairs (cf. section \ref{sec-4.4} and section \ref{sec-5}). Second, we will try to recruit more raters. For this purpose, however, we will not alter the rating task instructions - simply for the reason that all future raters will show the same error as the first five raters while rating 10 additional items.

\vfill

\pagebreak

\section{Appendix}
\label{appendix}
\begin{table}[h]
\caption{Examples for pairs likely not rated similar enough.}
\resizebox{0.75\textwidth}{!} {\begin{minipage}{\textwidth}
\renewcommand{\arraystretch}{1.5}
\centering
\begin{tabular}{p{.5cm}p{1.25cm}p{1.5cm}p{11cm}p{2cm}}\hline
Ex. (no.) & Entry ID & Term & Definition & Source\\ \hline
\multirow{2}{*}{1} & 916 & hazardous event & event that may result in harm & IEC 61508-4:2010 \citep{61508-4.2010} \\ \cline{2-5}
 & 1072 & incident & event that is not part of the expected operation of a system or service that causes, or may cause, an interruption to, or a reduction in, the quality of the service provided by the control system & IEC 62443-3-3:2013 \citep{62443-3-3.2013}\\ \hline
\multirow{2}{*}{2} & 1177 & denial of service & prevention or interruption of authorized access to a system resource or the delaying of system operations and functions & IEC TS 62443-1-1:2009 \citep{62443-1-1.2009}\\ \cline{2-5}
 & 1109 & incident & event that is not part of the expected operation of a system or service that causes or may cause, an interruption to, or a reduction in, the quality of the service provided by the system & IEC 62443-2-1:2010 \citep{62443-2-1.2010}\\ \hline
 \multirow{2}{*}{3} & 1190 & guard & gateway that is interposed between two networks (or computers or other information systems) operating at different security levels (one network is usually more secure than the other) and is trusted to mediate all information transfers between the two networks, either to ensure that no sensitive information from the more secure network is disclosed to the less secure network, or to protect the integrity of data on the more secure network & \multirow{2}{2cm}{IEC TS 62443-1-1:2009 \citep{62443-1-1.2009}} \\ \cline{2-4} 
 & 1188 & gateway & relay mechanism that attaches to two (or more) computer networks that have similar functions but dissimilar implementations and that enables host computers on one network to communicate with hosts on the other & \\ \hline
\end{tabular}
\end{minipage} }
\end{table}


\begin{table}[h]
\caption{Examples for pairs likely rated too similar.}
\resizebox{0.75\textwidth}{!} {\begin{minipage}{\textwidth}
\renewcommand{\arraystretch}{1.5}
\centering
\begin{tabular}{p{.5cm}p{1.25cm}p{1.5cm}p{11cm}p{2cm}}\hline
Ex. (no.) & Entry ID & Term & Definition & Source\\ \hline
\multirow{2}{*}{1} & 1198 & intrusion & unauthorized act of compromising a system & \multirow{2}{2cm}{IEC TS 62443-1-1:2009 \citep{62443-1-1.2009}}\\ \cline{2-4}
 & 1199 & intrusion detection & security service that monitors and analyzes system events for the purpose of finding, and providing real-time or near real-time warning of, attempts to access system resources in an unauthorized manner & \\ \hline
\multirow{2}{*}{2} & 1293 & decryption & process of changing ciphertext into plaintext using a cryptographic algorithm and key  & \multirow{2}{2cm}{IEC TR 62443-3-1:2009 \citep{62443-3-1.2009}}\\ \cline{2-4}
 & 1330 & symmetric key & single cryptographic key that is used with a secret (symmetric) key algorithm & \\ \hline
 \multirow{2}{*}{3} & 1195 & integrity & quality of a system reflecting the logical correctness and reliability of the operating system, the logical completeness of the hardware and software implementing the protection mechanisms, and the consistency of the data structures and occurrence of the stored data
  & IEC TS 62443-1-1:2009 \citep{62443-1-1.2009} \\ \cline{2-5}
 & 1105 & device requirements & countermeasure characteristics necessary for the devices within a zone to achieve the desired target security level & IEC 62443-2-1:2010 \citep{62443-2-1.2010}\\ \hline
\end{tabular}
\end{minipage} }
\end{table}

\begin{acknowledgments}
The authors gratefully acknowledge the support of the German Federal Ministry for Economic Affairs and Energy (BMWi) (Grants: 03TNG006A and 03TNG006B).
\end{acknowledgments}

\starttwocolumn
\bibliography{compling_style}
\end{document}